\documentclass[twocolumn,showpacs,preprintnumbers,amsmath,amssymb,nofootinbib]{revtex4}

\usepackage{graphicx}
\usepackage{dcolumn}
\usepackage{bm}

\begin{document}

%\preprint{UAB--FT--526}

\title{Non-Commutative Mechanics as a modification of space-time }

\author{C.A. Vaquera A. and J.L. Lucio M.}
\email{vaquera@fisica.ugto.mx,  lucio@fisica.ugto.mx}
\affiliation{Instituto de F\'{\i}sica, Universidad de Guanajuato,
Lomas del Bosque \# 103, Lomas del Campestre, 37150 Le\'on,
Guanajuato, Mexico. Tel/Fax. +52 477 7885100 }

\date{\today}

\begin{abstract}

\noindent We formulate non-relativistic classical and quantum
mechanics in the non-commutative two dimensional plane. The approach
we use is based on the Galilei group, where the non-commutativity is
seen as a central extension upon identification of the boost
generators with the position operator. We perform a systematic study
of the free particle, defined by the symmetries of the space time,
which include the no-commutativity. The symmetries at the classical
level are analyzed in terms of Noether's theorem. Canonical
quantization is presented and the representation of the
corresponding Heisenberg algebra is obtained. The path integral
representation and Wigner distribution function in phase space are
also discussed. We work out, both at the classical and at the
quantum level, the harmonic oscillator, avoiding the use of the
conventional non-canonical transformation that leads to a momentum
dependent potential. We use Einsteins' model for a solid to
corroborate that, according with intuition,  as a consequence of the
space fuzziness the entropy is a growing function of $\theta$ in the low
temperature regime.
\end{abstract}

\pacs{03.65.Ca, 03.65-w}% PACS, the Physics and Astronomy
\maketitle

\section{\label{intro}Introduction}

\noindent Non-commutativity (NC)\footnote{In the following we will
use NC to stand for non-commutative and M, C and Q for mechanics
classical and quantum respectively.} %$^1$%
can be interpreted either, as
an intrinsic property of space time \cite{connes}, or as a
low-energy consequence of the fundamental string theory
\cite{witten}.  In both cases, Non commutative quantum mechanics
(NCQM) is seen as a laboratory where the properties of NC field
theories can be studied. This has however lead to ambiguities of
what is meant by NC mechanics, in fact different points of view
exist in the literature which seem to be equally relevant. Among the
more frequently used are the one based on the Moyal product, {\it
i.e.} assume that the dynamics is described by the conventional
equations but using everywhere Moyal products, and the second
approach where one suppose that the only way NC enters is through
the Poisson Brackets (commutators in QM) of the position operators.
How is the NC related to the space-time, {\it i.e.} can we identify
the non-commuting operators with the position? and if so, is the
resulting theory consistent, do we have precise rules to describe
the physics in different reference frames?. From our view point one
should clearly state the assumptions made to define the NC mechanics
we use, in particular, one should know what is the effect of NC on a
free particle.
\bigskip

\noindent In this paper we adopt the first view, namely we assume NC
is intrinsic property of space time and require that NC be
incorporated in a consistent way. Indeed, an important property of
any formulation of the mechanics is the symmetry of the underlying
space-time, {\it i.e.} the relativity group on which it is based. In
fact, we will follow the procedure used to formulate field theory
where the free particle is treated using the symmetries of the space
time and then, once a consistent framework is available,
interactions are introduced using further arguments
(renormalizability, gauge invariance, SUSY). Thus, we first consider
in detail the free particle taking as basis a Hamiltonian
description based on a symplectic structure -- determined by the
symmetries of the NC space -- which we assume is the Galilei group.
It is important to remark that already at this level we have some
constraints since consistency of NCM with Galilei group requires:
\begin{itemize}
\item to work in the two dimensions, \item although
$\left\{x_i,x_j\right\} \neq 0$ necessarily
$\left\{p_i,p_j\right\} = 0$.
\end{itemize}
\bigskip

\noindent An important point to remark is that in our approach,
the Poisson Brackets (PB) (or commutators in the quantum theory)
of the coordinates is always non zero, we avoid any non-canonical
transformation leading to vanishing PB, otherwise it would be
inconsistent with the assumed Galilei algebra. Thus, for us, the
interaction of a charged particle in two dimensions with a
perpendicular constant magnetic field is not a typical example of
NC since in such a case the free particle reduces to the standard
commutative problem which can be consistently quantized and, after
that, the interaction can be introduced through minimal coupling.
We will comment on these points below, but we first refer to some
of the existing work in the literature. Since there are so many
publications in this area, we will restraint to those of direct
interest to our approach, and still, apologize for undeliberate
omissions.
\bigskip

\noindent From our point of view the formulation of NCM that not
only is consistent with, but actually is based on, the Galilei
algebra is of particular interest. In such an approach -- which is
valid only in two dimensions \cite{luki,levy} -- the NC appears,
together with the mass, as two central extensions of the algebra.
This was recognized by authors in \cite{luki} who build a model that
provides a realization of the symmetry. They also discuss the
quantization of the model using  the formalism of constraints and a
lagrangian including higher derivatives. In spite of the
completeness of the work in \cite{luki}, there are several topics
that deserve further discussion. Thus, in the present paper we make
a summary of non-commutative classical mechanics NCCM in two
dimensions. Besides the Hamiltonian formalism, the first order
Lagrangian formulation of NCCM and the relation between the
Hamiltonian and Lagrangian are also discussed following the approach
in \cite{janb,janbb}. The equivalence between those formalism
naturally leads to the conclusion that two dimensional NCCM can
always be treated as a system with second class constraints, at
least if interaction with gauge fields are not introduced
\cite{jackiw}. We discuss Noether's theorem and starting from the
symmetries \textbf{we derive} the Hamiltonian for the free particle
on the NC plane. Although this appears to be a trivial and long
exercise, in fact it shed some light on some of the questions
previously formulated. Indeed, besides helping in the identification
of the Hamiltonian (there is not consensus on this point, for
example according to the approach in \cite{free} the dynamics of a
free particle on a NC space is equivalent to the study of the
dynamics of this particle with charge $q$ on the usual commutative
space in the presence of a magnetic field), the symmetries
determines the Hamiltonian as the time evolution operator, providing
thus some support to the whole approach. All along the the paper we
emphasize the role played by symmetries, this is relevant since the
proper formulation and solution of some of the problems in QM can be
traced to the appropriated understanding of the classical
symmetries.
\bigskip

\noindent With the formulation of the mechanics previously discussed
the passage from classical to quantum mechanics is straightforward.
At the quantum level, using canonical quantization,  we analyze the
representation of the Heisenberg algebra as a way to implement the
canonical quantization. As explained below, the idea is to implement
representations for the operators which are consistent with the
commutation relations, \textbf{but also} determine what is the
relevance of the commutation relation for the states. The advantage
of this approach is that, based on the commutation relations, one
can implement and compare different representations. An example
where a judicious choice of the base is exploited is  seen in
\cite{lumbo} where the periodicity of the oscillator in the
non-commutative plane is analyzed, although neither the spectrum nor
the eigenfunctions of the complete set of observables  are derived.
Previous work include also an analysis of possible realizations of
the operators, which is a related but not equivalent problem
\cite{polyn}, \cite{brihaye} . We derive expression for
non-equivalent representations of the Heisenberg algebra extending
previous analysis to NCQM \cite{janv}. The representations include
differential representations of the operators (position and momenta)
in different basis as well as gauge fields that follow from the
structure of the algebra and that are relevant in the description of
non trivial manifolds. As far as we know, ours is the first time a
detailed analysis of the representations of the Heisenberg algebra
is presented in the two dimensional NC case \cite{aci}.
\bigskip

\noindent An alternative procedure to study the quantum properties
of a system is through the phase space Wigner distribution
function. Besides the intrinsic interest on this formulation, the
use of non-commutativity in the field of quantum optics
\cite{op1}, where Wigner distribution function is a common tool,
is a further motivation for its generalization to the
non-commutative case. We are not aware of exhaustive work along
this line, the difference with existing literature being again the
approach \cite{koka}. We provide a compact exact expression for
the Wigner distribution function, again in this case we do not
need  to consider an expansion in $\theta$, the NC parameter. For
completeness we include a brief discussion on the path integral
formulation, where the results of the canonical quantization are
used, in particular the wave functions that permit to connect
different basis. Our results are analogous to those obtained in
\cite{aci}.
\bigskip

\noindent As examples of the application of the formalism, we work
out in detail two problems, both at the classical and at the quantum
level: the free particle and the harmonic oscillator. The free
particle can be considered the physical system where the symmetries
of the space time are realized, for that reason we characterize  the
free particle with the Hamiltonian that is compatible with the
corresponding Galilei group. On the other hand, several papers
\cite{osc1,osc2,su(2)} have dealt with the harmonic oscillator at
the quantum level. The case of the two dimensional harmonic
oscillator in the presence of a constant magnetic field has been
considered \cite{osc1,osc2} and it has been remarked that two phases
exist, in one of these ($\textbf{B} \neq 0$) the symmetry group is
$SU(2)$ while in the other ($\textbf{B} = 0$) the symmetry group is
$SU(1,1)$. In our approach we show in CM, that although SU(2) is a
symmetry of the commutative harmonic oscillator, this is not true in
the NC case. We analyze the surviving symmetry and, at the quantum
level, determine the eigenfunctions common to the Hamiltonian and
angular momentum.
\bigskip

\noindent Before concluding this introduction in this paragraph we
make a summary of our approach. We consider the mechanics, both
classical and quantum, for a free particle imposing consistency with
the symmetry of the space time, which is assumed to be Galilei group
including two central extensions (the mass and the parameter of NC)
. We completely avoid the use of non canonical transformations and
assume that the only way NC enters is through the commutation
relations of the position operators plus those that the requirement
of consistency demands. A potential $V(x)$,  independent of the NC
parameter $\theta$, is added to the Hamiltonian of the free
particle, which has been previously determined through the symmetry.
In QM we assume the validity of the Schr\"{o}dinger equation and use
the time translation generator, {\it i.e.} the Hamiltonian. The
differential operator associated to the Hamiltonian is obtained from
the classical one plus any of the four representations of the
position and momenta consistent with the Heisenberg algebra. We have
tried to make the paper as self contained as possible, avoiding
however unnecessary details. We organized the main body of the
manuscript in six parts. Section two is devoted to classical
mechanics, the third to the canonical quantization. Sections four
and five deal with alternative quantization procedures. Examples in
classical mechanics are discussed in section two, in QM in section
six and we end with a summary of the contributions of this work.

\section{Classical Mechanics}

\noindent We start with a brief summary of classical mechanics. This
will allow us to introduce the notation and also to clarify the role
played by symmetries in the formulation of NC mechanics. In the
Hamiltonian formalism the description of a system with $n$ degrees
of freedom is determined by $2n$ first order differential equations
involving $2n$ independent variables that, here and thereof, we
denote by $z_\alpha$ $(\alpha = 1,2, ... 2n)$. The symplectic
structure $J$ associated to this formalism defines the Poisson
brackets
\begin{equation}
\label{parepoi}
\left\{  f,g\right\}  =J^{\alpha\beta}\frac{\partial f}{\partial z^{\alpha}%
}\frac{\partial g}{\partial z^{\beta}}~,%
\end{equation}
\noindent where summation over repeated indices is understood. The
parenthesis are real, antisymmetric and linear. Furthermore they
satisfy the Leibiniz rule and the Jacobi identity:
\begin{itemize}
\item$\{f,gh\}=\{f,g\}h+g\{f,h\}$ \\
\item$\{f,\{g,h\}\}=\{\{f,g\},h\}+\{g,\{f,h\}\}$.
\end{itemize}
\noindent According to (\ref{parepoi}) the entries of $J$ are:
\begin{equation}
\ J^{\alpha\beta}=-J^{\beta\alpha}=\left\{
z^{\alpha},z^{\beta}\right\}  ,
\end{equation}
\noindent For later reference we introduce the inverse of the
symplectic structure $J$  by means of the relation:
\begin{equation}
J^{\alpha\beta}\omega_{\beta\gamma}=\delta_{\gamma}^{\alpha},
\end{equation}
\noindent The equations of motion for a system described by the
Hamiltonian $H(z)$, in a phase space with symplectic structure
$J$, are given by:
\begin{equation}
\dot{z}^{\alpha}=\left\{  z^{\alpha},H(z)\right\}  =J^{\alpha\beta}%
\frac{\partial H(z)}{\partial z^{\beta}},\text{\ \ \ }\alpha,\beta
=1,\ldots,2n. \label{ecmov}%
\end{equation}
If one identifies the generalized coordinates with the first $n$
variables ($q_i=z_i, i=1,2, ... n$) of phase space and the
conjugated momenta with the last $n$ ($p_i=z_i, i=n+1,n+2, ...
2n$), then the information on the symplectic structure $J$ is
contained in the Poisson brackets:
\begin{equation}%
\begin{array}
[c]{ccc}%
\{q_{\alpha},q_{\beta}\}=\theta_{\alpha\beta}, &
\{q_{\alpha},p_{\beta}\}=\delta_{\alpha\beta}, & \{p_{\alpha}%
,p_{\beta}\}=\Theta_{\alpha\beta},
\end{array}
\label{conm1}%
\end{equation}
\noindent So far $\theta_{\alpha\beta}=\theta_{\alpha\beta}(q,p)$ is
an antisymmetric field ($\theta
_{\alpha\beta}=-\theta_{\beta\alpha}$) which, in general, may depend
upon the position and the momenta. However, if we set
$\Theta_{\alpha\beta}=0$, the Jacobi identity (applied to a momentum
$p_i$ and two coordinates  $q_j,q_k$), implies the relation:
\begin{equation}
\frac{\partial\theta_{ij}}{\partial q_{k}}=0,
\end{equation}
\noindent therefore $\theta_{\alpha\beta}=\theta_{\alpha\beta} (p)$.
\noindent At this point it should be clear that, by allowing a
general enough symplectic structure,  NCM can be described within
this formalism.
\bigskip

\noindent On the other hand, the lagrangian formulation is
relevant in discussing the symmetries. For that reason we
introduce  the first order Lagrangian \cite{janb}:
\begin{equation}
L(z,\dot{z},t)=K_{\alpha}(z)\dot{z}^{\alpha}-H(z,t),\text{ \ \ \
}\left( \alpha=1,\cdots,2n\right)
\end{equation}
\noindent {where} $K_\alpha$ is a vector potential in phase space
for the inverse of the symplectic structure, {\it i.e.}:
\begin{equation}
\omega_{\alpha\beta}=\frac{\partial K_{\beta}}{\partial z^{\alpha}}%
-\frac{\partial K_{\alpha}}{\partial z^{\beta}};
\end{equation}
\noindent in order to show the equivalence of this Lagrangian to
the Hamiltonian formalism previously introduced, we apply the
variational principle to the action associated to this Lagrangian:
\begin{align}
&  \delta S[z(t)]
 =\delta\int_{t_{i}}^{t_{f}}\left(K_{\alpha}(z)\dot{z}^{\alpha}-H(z,t)\right) dt\label{varia}\\
&  =\int_{t_{i}}^{t_{f}}\left(  \frac{\partial K_{\alpha}}{\partial z^{\beta}%
}\dot{z}^{\alpha}\delta z^{\beta}+K_{\alpha}\frac{d}{dt}\delta
z^{\alpha
}-\frac{\partial H}{\partial z^{\beta}}\delta z^{\beta}\right)  dt\nonumber\\
&  =\int_{t_{i}}^{t_{f}}\left[  \omega_{\beta\alpha}\dot{z}^{\alpha}%
-\frac{\partial H}{\partial z^{\beta}}\right]  \delta
z^{\beta}dt+(K_{\alpha }\delta
z^{\alpha})|_{t_{i}}^{t_{f}}.\nonumber
\end{align}
\noindent Therefore, if the variation at the end points are such
that:
\begin{equation}
K_{\alpha}\delta z^{\alpha}|_{t_{i}}=K_{\alpha}\delta
z^{\alpha}|_{t_{f}},
\end{equation}
\noindent the variation of the action Eq.(\ref{varia}) implies:
\begin{equation}
\omega_{\beta\alpha}\dot{z}^{\alpha}=\frac{\partial H}{\partial
z^{\beta}},
\end{equation}
which are nothing but the Hamilton equations of motion
Eq.(\ref{e1}):
\begin{equation}
\dot{z}^{\alpha}=\left\{  z^{\alpha},H(z)\right\}  =J^{\alpha\beta}%
\frac{\partial H(z)}{\partial z^{\beta}},
\end{equation}
\noindent If $\omega(z)$ has constant entries, $K_\alpha$ can be
written as:
\begin{equation}
K_{\alpha}=\frac{1}{2}z^{\beta}\omega_{\beta\alpha}.
\end{equation}
\noindent and, under these conditions, the Lagrangian reduces to:
\begin{equation}
L(z,\dot{z},t)=\frac{1}{2}z^{\beta}\omega_{\beta\alpha}\dot{z}^{\alpha
}-H(z,t),\text{ \ \ }%
\end{equation}
\noindent With the lagrangian formulation at hand, Noether's
theorem is formulated in the conventional way, for completeness we
quote the result \cite{janb}.
\bigskip

\noindent If under a group of order $r$, of continuous
transformations $t \to t'=t'(t), z'^\alpha \to z'^\alpha(z,t)$,
the action is invariant up to surface terms {\it i.e.}
\begin{equation}
S[z^{\prime}]=S[z]+\int_{t_{i}}^{t_{f}}\frac{d\Lambda(z)}{dt}dt.
\end{equation}
\noindent then, for every classical solution to the equations of
motion, $r$ functions $\gamma_\lambda$ $(\lambda = 1,2, ...r)$  of
the dynamical variables are conserved:
\begin{equation}
\gamma_{\lambda}=K_{\alpha}\varphi_{\lambda}^{\alpha}-\chi_{\lambda}\left(
H-\dot{z}^{\alpha}K_{\alpha}\right)  -\Lambda_{\lambda}. \label{noch}%
\end{equation}
\noindent the quantities $\phi,\chi$ and $\Lambda$ are given by
the infinitesimal transformations which are a symmetry of the
action:
\begin{align}
\delta t  &  =\varepsilon^{\lambda}\chi_{\lambda}(t),\label{no1}\\
\delta z^{\alpha}  &
=\varepsilon^{\lambda}\varphi_{\lambda}^{\alpha
}(z),\nonumber\\
\delta\Lambda &  =\varepsilon^{\lambda}\Lambda_{\lambda},\nonumber
\end{align}
\noindent {where} $\varepsilon^{\lambda}$ ($\lambda=1,\cdots,r$)
are constant, infinitesimal parameters each of which is associated
to a given transformation. It proofs convenient to introduce the
notation $Q= \varepsilon^{\lambda}\gamma_\lambda$ (no sum over
$\lambda$) and refer to $Q$ as the charge associated to the
corresponding transformation.
\bigskip

\noindent It is well known  \cite{?gold} that the Poisson brackets
of the conserved charges $\gamma_\lambda$ define an algebra
isomorphic to the global continuous symmetry group of the
Lagrangian, and that the symmetry transformations can be obtained in
terms of the Poisson brackets of the dynamical variables
($z_\alpha$) with $\gamma_\lambda$. Instead of checking the validity
of this assertion, later in this section we will use this fact to
obtain the generators and, from these, the hamiltonian for a free
particle.
\bigskip

\noindent This is as far as we can get on general grounds. To go
further we need to specify the symplectic structure. We do this
taking into account facts known from the mathematical literature
regarding the Galilei group \cite{gal}, which we assume is the
symmetry group of the non-relativistic mechanics. It is known that
in $3+1$ dimensions, the Galilei group accepts only one central
charge - to be identified with the mass of the particle - and
therefore it is not possible to introduce NC in 3+1 dimensions, at
least not consistently with the Galilei group. In $2+1$ dimensions
the Galilei algebra accepts three central extensions: the mass,
the parameter associated to NC and one more that we ignore on
physical grounds (we are not interested in such an extension).
Note in particular that consistency with Galilei group demands the
vanishing of the Poisson Bracket among the momenta\footnote{At this point it is not clear the relation between the
Galilei algebra and the symplectic structure. The connection
between these structures is seen when the generators of the
Galilei group are expressed in terms of the dynamical variables of
a physical system, see below.}%$^2$% 
Thus, we
restraint our analysis to $2+1$ dimensions, and assume the
symplectic structure given by Eq.(\ref{conm1}), where now
$\Theta=0$ and we further assume that:

\begin{equation}
\theta_{ij}=\theta\epsilon_{ij}, \label{th1}%
\end{equation}
\noindent with $\theta$ a constant parameter, which clearly
characterize the non-commutativity. The Poisson brackets (PB) and
the equations of motion read respectively:
\begin{equation}
\left\{  f,g\right\}  =\frac{\partial f}{\partial
q_{i}}\frac{\partial g}{\partial p_{i}}-\frac{\partial f}{\partial
p_{i}}\frac{\partial g}{\partial
q_{i}}+\theta\epsilon_{ij}\frac{\partial f}{\partial
q_{i}}\frac{\partial
g}{\partial q_{j}}, \label{c1}%
\end{equation}
\begin{align}
\dot{q}_{i}  &  =\frac{\partial H}{\partial
p_{^{_{i}}}}+\theta\epsilon
_{ij}\frac{\partial H}{\partial q_{j}},\label{e1}\\
\dot{p}_{_{^{i}}}  &  =-\frac{\partial H}{\partial
q_{i}},\nonumber
\end{align}
\noindent Let us consider now a system invariant under spatial
instantaneous transformations:
\begin{equation}
\delta t =0,~~~~~~ \delta q_{i}   =b_{i},~~~~~~ \delta p_{i}  =0,
\end{equation}
\noindent where $b_{i}$ ($i=1,2$) are arbitrary infinitesimal
parameters. In order to identify the generators of this
transformations we write:
\begin{align}
\delta q_{i}  &  =\{q_{i},Q_{trans}\}=\{q_{i},b_{j}\gamma_{j}\}=b_{i}, \label{detrans}\\
\delta p_{i}  &
=\{p_{i},Q_{trans}\}=\{p_{i},b_{j}\gamma_{j}\}=0.\nonumber
\end{align}
\noindent Thus, using (\ref{c1}) we conclude that the space
translation generators satisfy the equations:
\begin{align}
\frac{\partial\gamma_{j}}{\partial p_{i}}+\theta\epsilon_{ik}\frac
{\partial\gamma_{j}}{\partial q_{k}}  &  =\delta_{ij},\\
-\frac{\partial\gamma_{j}}{\partial q_{k}}  &  =0,\nonumber
\end{align}
\noindent and then, up to a constant, the space translation
generators are the momentum components $\gamma_{i}=p_{i}.$
\noindent In a similar way we can treat the boost transformations:
\begin{equation}
\delta t  =0,~~~~~ \delta q_{i} =v_{i}t,~~~~~ \delta p_{i}
=mv_{i}.
\end{equation}
\noindent {where} $v_{i}$ ($i=1,2$) are the infinitesimal
parameters associated to  boosts. In this case the transformation
of the position and momentum are given by:
\begin{align}
\delta q_{i}  &  =\{q_{i},Q_{boost}\}=\{q_{i},-k_{j}v_{j}\}=v_{i}t,\\
\delta p_{i}  &
=\{p_{i},Q_{boost}\}=\{p_{i},-k_{j}v_{j}\}=mv_{i}.\nonumber
\label{deboost}
\end{align}
\noindent Using (\ref{c1}) and (\ref{detrans}) Eq.(\ref{deboost})
reduces to:
\begin{align}
\frac{\partial k_{j}}{\partial p_{i}}+\theta\epsilon_{ik}\frac{\partial k_{j}%
}{\partial q_{k}}  &  =-t\delta_{ij},\\
\frac{\partial k_{j}}{\partial q_{k}}  &  =m\delta_{jk},\nonumber
\end{align}
therefore the \textsl{boosts} generators $k_{i}$ are:
\begin{equation}
k_{i}=mq_{i}-p_{i}t+m\theta\epsilon_{ij}p_{j}, \label{boost}%
\end{equation}
\noindent Likewise, we obtain for the angular momentum, the
generator of spatial rotations:
\begin{equation}
J=\epsilon_{ij}q_{i}p_{j}+\frac{\theta}{2}p_{k}p_{k}. \label{MomAng}%
\end{equation}
\noindent We can now determine the Hamiltonian of a system
possessing all the symmetries of the Galilean group, which we take
as the definition of a free particle. To this end we consider the
time derivative of an arbitrary function of the phase space
variables:
\begin{equation}
\frac{df}{dt}=\frac{\partial f}{\partial t}+\frac{\partial
f}{\partial z^{\alpha}}\dot{z}^{\alpha}.
\end{equation}
\noindent which by using the equations of motion (\ref{ecmov}), is
cast in the form:
\begin{equation}
\frac{df}{dt}  =\frac{\partial f}{\partial t}+\left\{
f,H(z)\right\} .\nonumber
\end{equation}
\noindent Applying this relation to Noether's charges ($\dot
Q=0$)we obtain
 the following relations:
\begin{eqnarray}
\left\{  H,H\right\} =0,~~~~~~~\left\{  p_{i},H\right\} =0,\\
\left\{  J,H\right\} =0,~~~~~~ \left\{ k_{i},H\right\} =p_{i}
\label{plib} \nonumber.
\end{eqnarray}
\noindent The Poisson brackets Eq.(\ref{plib}), after using
(\ref{c1}), (\ref{boost}) and (\ref{MomAng}), amounts to the set
of simultaneous equations:
\begin{equation}
\frac{\partial H}{\partial q_{i}} =0,~~~~~~~~~~~ \frac{\partial
H}{\partial p_{i}}  =\frac{p_{i}}{m}.\nonumber
\end{equation}
\noindent Thus, we conclude that the Hamiltonian for a free
particle of mass $m $, in a non-commutative plane, is given by:
\begin{equation}
H=\frac{p_{i}p_{i}}{2m} \label{hfp},
\end{equation}
\noindent This is a complicated form of deriving the conventional
Hamiltonian for a free particle, in this way however we are
certain that the description is consistent with the Galilean
relativity. Furthermore, it is important to remark that for a
explicitly time-independent Hamiltonian $H=H(z)$, the following
transformation always define a global invariance of the system
\begin{align}
\delta z^{\alpha}  &  =-\frac{d z^{\alpha}}{dt}\delta t,\nonumber\\
\delta t  &  =\tau,
\end{align}
where $\tau$ is a constant parameter. \ Therefore, the generator
of time translations is simply
\begin{equation}
\gamma=-H-\Lambda.
\end{equation}
Therefore, for every system described by a time-independent
Hamiltonian, the Hamiltonian itself is a constant of motion.

\bigskip

\noindent For completeness we quote the full Galilei algebra, which
is obtained through the Poisson brackets and the expressions for the
generators previously derived:
\begin{equation}
\begin{array}
[c]{cc}%
\left\{  p_{i},H\right\}  =0, & \left\{  p_{i},p_{j}\right\}  =0,\\
\left\{  J,H\right\}  =0, & \left\{  J,p_{i}\right\}  =\epsilon_{ij}p_{j},\\
\left\{  k_{j},H\right\}  =p_{j}, & \left\{  k_{j},p_{i}\right\}
=m\delta_{ji},\\
\left\{  J,k_{i}\right\}  =\epsilon_{ij}k_{j}, & \left\{
k_{i},k_{j}\right\} =-m^{2}\theta\epsilon_{ij}.
\end{array}
  \label{algal}%
\end{equation}
\noindent The last two relations show the appearance of the central
extensions, the mass $m$ and the NC parameter $\theta$. Thus, {\bf
the symplectic structure used to describe the mechanics follows from
the symmetry of the space-time described by the Galilei algebra}.
\bigskip

\noindent We now consider the examples at the classical level
 \cite{classical}. To start let us consider the free particle. The
Hamiltonian Eq.(\ref{hfp}) leads to the equations:
\begin{equation}
\dot{q}_{i} =\frac{p_{i}}{m},~~~~~~~~~~~~\dot{p}_{i}=0.\nonumber
\end{equation}
The solution is the the conventional one:
\begin{equation}
q_{i} =\dot{q}_{i0}t+q_{i0}, ~~~~~~~p_{i}=m \dot{q}_{i0},\nonumber
\end{equation}
\noindent with $q_{i}(0)=q_{i0}$ and $\dot{q}_{i}(0)=\dot{q}_{i0}$
given initial conditions. A more involved problem is the system,
to which, for obvious reasons we will refer as the harmonic
oscillator, characterized by the Hamiltonian,:
\begin{equation}
H=\frac{1}{2m}p_{i}p_{i}+\frac{m\omega^{2}}{2}q_{i}q_{i}.
\end{equation}
In this case the relations arising from the Hamiltonian formalism
can be combined to produce the following equations for $x=q_{1}$
and $y=q_{2}$:
\begin{align}
\ddot{x}  &  =-\omega^{2}x+m\theta\omega^{2}\dot{y}\\
\ddot{y}  &  =-\omega^{2}y-m\theta\omega^{2}\dot{x}.\nonumber
\end{align}
Denoting the initial conditions by $x(0)=x_{0}$,
$\dot{x}(0)=\dot{x}_{0}$, $y(0)=y_{0}$ y $\dot
{y}(0)=\dot{y}_{0}$, the solution to the harmonic oscillator is
given by;
\begin{align}
x(t)    = ~& ~T_1(t) x_{0} ~+ ~T_2(t) \dot{x}_{0} \label{sol1}
  +~ T_3(t) \left(  m\theta\omega^{2}x_{0}+2\dot{y}
_{0}\right) \nonumber \\
   & - T_4(t) \left( m\theta\dot{x}_{0}+2y_{0}\right) \\
y(t)   =~ & T_1(t) y_{0}~+~ T_2(t)\dot{y}_{0}\label{sol2} +~
T_3(t)  \left(  m\theta\omega^{2}y_{0}-2\dot{x}
_{0}\right)  \nonumber\\
& - T_4(t) \left( m\theta\dot{y}_{0}-2x_{0}\right)\nonumber
\end{align}

\noindent where we have introduced the following definitions:
\begin{equation}
T_1(t)    =\frac{ \cos(\phi t)+\cos(\chi t) }{2},~ T_4(t)
=\frac{\phi\sin(\chi t)-\chi\sin(\phi
t)}{2\omega\sqrt{4+m^{2}\omega^{2}\theta^{2}}}\nonumber
\end{equation}
\noindent
\begin{equation}
T_2(t)   = \frac{\phi\sin(\chi t)+\chi\sin(\phi t)
}{2\omega^{2}},~ T_3(t) =\frac{\cos(\chi t)-\cos(\phi
t)}{2\omega\sqrt{4+m^{2}\omega^{2}\theta^{2}}}\nonumber
\end{equation}
\begin{equation}
\phi, \chi  =\sqrt{\omega^{2}+\frac{1}{2}m^{2}\theta^{2}\omega^{4} \pm \frac{1}%
{2}m\theta\omega^{3}\sqrt{4+m^{2}\theta^{2}\omega^{2}}},\nonumber\\
\end{equation}

\noindent As expected, the solution to the NC case,
Eq.(\ref{sol1}) reduce to the conventional harmonic oscillator in
the $\theta\rightarrow0$ limit. In order to asses the effect of NC
on this system, it is convenient to compare the solution to the
commutative case. This is easier to do in the
$m\theta\omega^{2}<<\omega$, where one shows that:
\begin{equation}
x+iy\simeq(x^{(\theta=0)}+iy^{(\theta=0)})e^{-\frac{i}{2}(m\theta\omega^{2}%
t)}.
\end{equation}
\noindent and so we conclude that the effect of NC amounts to
rotate, in the $x-y$ plane, the commutative solutions with angular
velocity $m\theta\omega^{2}/2$.
\bigskip

\noindent To conclude this section we consider the symmetry group
of the harmonic oscillator. Since the conventional and NC cases
are different, and can not be obtained one from the other (see
below), for the sake of clarity we treat both of them. The
procedure we follow consist in parameterizing the conserved
quantities and then to obtain the parameters through the use of
the Jacobi identity. In fact, if the conserved quantities do not
depend explicitly of time, then the equation of motion implies:
\begin{equation}
\left\{  H,S_{i}\right\}  =0. \label{cmov1}%
\end{equation}
\noindent On the other hand, using the Jacobi identity we obtain:
\begin{equation}
\left\{  H,\left\{  S_{i},\cdot\right\}  \right\}  -\left\{
S_{i},\left\{ H,\cdot\right\}  \right\}  =\left\{  \left\{
H,S_{i}\right\}  ,\cdot\right\}
=0, \label{cmov2}%
\end{equation}
\noindent Let us begin with the commutative case. We will use a
subscript or superscript $_0$ to avoid confusion with the analogous
problem in the NC case. The PB of the Hamiltonian with an arbitrary
function is:
\begin{equation}
\left\{  H,\cdot\right\}
_{0}=-\frac{p_{x}}{m}\frac{\partial}{\partial
x}-\frac{p_{y}}{m}\frac{\partial}{\partial
y}+m\omega^{2}x\frac{\partial }{\partial
p_{x}}+m\omega^{2}y\frac{\partial}{\partial p_{y}}\label{hconm} .
\end{equation}

\noindent We parameterize the PB of the constant of motion $S_i^0$
with an arbitrary function in the following form:
\begin{equation}
\left\{  S_{i}^{0},\cdot\right\}  _{0}=\sum_{\alpha=1}^{4}\sum_{\beta=1}%
^{4}\left(  \left.  a_{(i)}^{0}\right.
_{\alpha\beta}z^{\beta}\right) \frac{\partial}{\partial
z^{\alpha}} \label{paramet},
\end{equation}
\noindent By using this parametrization we assume the conserved
quantities are bilinear in the phase space variables. Furthermore
notice that we have to invert and integrate Eq.(\ref{paramet}) to
obtain the $S_i^0$. The matrix elements $\left. a_{(i)}^{0}\right.
_{\alpha\beta}$ are constant, unknown parameters, to be determined
by the relations (\ref{cmov1}) y (\ref{cmov2}). After a lengthly
calculation one concludes that the relations give rise to the
following four linearly independent matrices
$\mathbf{a}_{(i)}^{0}$:
\begin{equation}%
\begin{array}
[c]{c}%
\mathbf{a}_{0}^{0}=\frac{1}{m}\left(
\begin{array}
[c]{cccc}%
0 & 0 & -1 & 0\\
0 & 0 & 0 & -1\\
m^{2}\omega^{2} & 0 & 0 & 0\\
0 & m^{2}\omega^{2} & 0 & 0
\end{array}
\right)  ,\\
\mathbf{a}_{1}^{0}=\frac{1}{2m\omega}\left(
\begin{array}
[c]{cccc}%
0 & 0 & 0 & -1\\
0 & 0 & -1 & 0\\
0 & m^{2}\omega^{2} & 0 & 0\\
m^{2}\omega^{2} & 0 & 0 & 0
\end{array}
\right)  ,\\
\mathbf{a}_{2}^{0}=\frac{1}{2}\left(
\begin{array}
[c]{cccc}%
0 & 1 & 0 & 0\\
-1 & 0 & 0 & 0\\
0 & 0 & 0 & 1\\
0 & 0 & -1 & 0
\end{array}
\right)  ,\\
\text{\ }\mathbf{a}_{3}^{0}=\frac{1}{2m\omega}\left(
\begin{array}
[c]{cccc}%
0 & 0 & 1 & 0\\
0 & 0 & 0 & -1\\
-m^{2}\omega^{2} & 0 & 0 & 0\\
0 & m^{2}\omega^{2} & 0 & 0
\end{array}
\right)  ,
\end{array}
\end{equation}
\noindent from which we obtain the conserved quantities:
\begin{align}
S_{0}^{0} &  =H, \label{symho}\\
S_{1}^{0} &  =\frac{1}{2m\omega}\left(
p_{x}p_{y}+m^{2}\omega^{2}xy\right)
,\nonumber\\
S_{2}^{0} &  =\frac{1}{4m\omega}\left[
p_{y}^{2}-p_{x}^{2}+m^{2}\omega
^{2}\left(  y^{2}-x^{2}\right)  \right]  ,\nonumber\\
S_{3}^{0} &  =\frac{J_{0}}{2}=\frac{1}{2}\left(
xp_{y}-yp_{x}\right) .\nonumber
\end{align}
\noindent It is clear that the four quantities are not
independent, in fact it is easy to show that the following
relation holds:
\begin{equation}
\left(  S_{1}^{0}\right)  ^{2}+\left(  S_{2}^{0}\right)
^{2}+\left( S_{3}^{0}\right)  ^{2}=\frac{H^{2}}{4\omega^{2}}.
\end{equation}
\noindent Moreover one verifies that the $S_i^0$ $(i=1,...,3)$
fulfill the following algebra:
\begin{equation}
\left\{  S_{i}^{0},S_{j}^{0}\right\}
_{0}=\epsilon_{ijk}S_{k}^{0}.
\end{equation}
Thus we have re-derived the well known result that $SU(2)$ is the
symmetry group of the two dimensional commutative harmonic
oscillator.
\bigskip

\noindent We now consider the symmetry group for the NC harmonic
oscillator \cite{cvtesis}. Instead of Eq.(\ref{hconm}) we have to
use:
\begin{align}
\left\{  H,\cdot\right\}   &  =-\left(  \frac{p_{x}}{m}+\theta
m\omega ^{2}y\right)  \frac{\partial}{\partial x}-\left(
\frac{p_{y}}{m}-\theta
m\omega^{2}x\right)  \frac{\partial}{\partial y}\label{cmov3}\\
&  +m\omega^{2}x\frac{\partial}{\partial
p_{x}}+m\omega^{2}y\frac{\partial }{\partial p_{y}}.\nonumber
\end{align}
\noindent  We assume again Eq.(\ref{paramet}) for the PB of the
conserved quantities. In this case the constraints due to Eqs.(
\ref{cmov1}, \ref{cmov2}) lead only  two linearly independent
matrices $\mathbf{a}_{(i)}$:
\begin{equation}%
\begin{array}
[c]{c}%
\mathbf{a}_{0}=\frac{1}{m}\left(
\begin{array}
[c]{cccc}%
0 & -\theta m^{2}\omega^{2} & -1 & 0\\
\theta m^{2}\omega^{2} & 0 & 0 & -1\\
m^{2}\omega^{2} & 0 & 0 & 0\\
0 & m^{2}\omega^{2} & 0 & 0
\end{array}
\right)  ,\label{matrixtheta}\\
\mathbf{a}_{1}=\left(
\begin{array}
[c]{cccc}%
0 & 1 & 0 & 0\\
-1 & 0 & 0 & 0\\
0 & 0 & 0 & 1\\
0 & 0 & -1 & 0
\end{array}
\right)  ,
\end{array}
\end{equation}
\noindent so that for the NC oscillator the constants of motion are:
\begin{align}
S_{0}  &  =H, \label{symtheta}\\
S_{1}  &  =J=xp_{y}-yp_{x}+\frac{\theta}{2}\left(
p_{x}^{2}+p_{y}^{2}\right) .\nonumber
\end{align}
\noindent We conclude that \textbf{$SU(2)$ is not} a symmetry of
the harmonic oscillator on the NC plane \cite{symmetries}. The
fact that diverse conclusions regarding the symmetry of the NC
armonic oscillator are reached by different authors
\cite{osc1,osc2,su(2)} is a manifestation of the ambiguities in
defining the theory. A final comment is in order. Comparing
Eqs.(\ref{symho},\ref{symtheta}) we observe that in the
$\theta\rightarrow0$ limit, the symmetry of the commutative
harmonic oscillator is not recovered. Operationally, it is clear
that in the $\theta\rightarrow0$ limit, Eqs.(\ref{matrixtheta},
\ref{symtheta}) reproduce only two of the three independent
conserved quantities.

\section{Quantum Mechanics}

\noindent In this section we discuss the problem of quantization
assuming that a Hamiltonian description -- as presented in the
previous section -- is available for the corresponding classical
system. The quantization proceeds through the correspondence
principle or \textsl{canonical quantization} \cite{Dirac} which
associates to the classical phase space a quantum space of states
described in terms of a Hilbert space. To the fundamental degrees
of freedom $z^{\alpha}$ the principle associates linear operators
$\hat{z}^{\alpha}$, which act on the Hilbert space. The
commutation relations of the quantum operators are obtained
multiplying the classical PB by $i\hbar$ \footnote{The $\hbar$ enters naturally in the quantization procedure,
just on dimensional grounds one requires it. In the NC case the
situation is different, in fact one should keep in mind the
appropriated units of $\theta$, which are not the same that at the
classical level.}.%$^3$

\begin{equation}
\{z^{\alpha},z^{\beta}\}=J^{\alpha\beta} \to \left[
\hat{z}^{\alpha},\hat{z}^{\beta}\right]  =i\hbar J^{\alpha\beta},
\end{equation}
\noindent Very much as the PB define a geometric structure on
phase space, the Hilbert space possess an algebraic structure in
the sense that it provides a linear representation of the algebra
of the quantum operators \cite{quien}.
\bigskip

\noindent The internal product of the Hilbert space must satisfy
two requirements. First the operators associated to the classical
fundamental degrees of freedom must be Hermitian and self-adjoint
respect to the internal product, and second the internal product
must be Hermitian, {\it i.e.}
\begin{equation}
\left\langle \psi|\phi\right\rangle ^{\ast}=\left\langle \phi|\psi
\right\rangle ,
\end{equation}
the * stands for complex conjugation and $\left| \psi\right\rangle
$, $\left|  \phi\right\rangle $ are arbitrary quantum states.
\bigskip

\noindent Thus, the description of the dynamics of a quantum
system use the same structures as the classical one, namely the
Hamiltonian and the symplectic structure. The time evolution of
the system is given by the Schr\"{o}dinger equation:
\begin{equation}
\hat{H}\left|  \psi;t\right\rangle =i\hbar\frac{d}{dt}\left|  \psi
;t\right\rangle , \label{Schr}%
\end{equation}
where $\hat{H}$ is the quantum Hamiltonian (Hermitian,
self-adjoint, obtained from the classical time translation
generator) and $\left| \psi;t\right\rangle $ the state of the
system at time $t$.
\bigskip

\noindent From now on we use the notation
$x=q_{1},y=q_{2},p_{x}=p_{1}$ and $p_{y}=p_{2}$. The commutation
relations are then:
\begin{equation}
\left\{
\begin{array}
[c]{c}%
\lbrack\hat{x},\hat{x}]=[\hat{y},\hat{y}]=[\hat{x},\hat{p}_{y}]=[\hat{y}%
,\hat{p}_{x}]=0,\text{ \ }\\
\lbrack\hat{p}_{x},\hat{p}_{x}]=[\hat{p}_{y},\hat{p}_{y}]=[\hat{p}_{x},\hat
{p}_{y}]=0\text{\ }\\
\lbrack\hat{x},\hat{y}]=\text{\ }i\hbar\theta,\text{ \ \ \ \
}[\hat{x},\hat
{p}_{x}]=[\hat{y},\hat{p}_{y}]=i\hbar.\text{ \ }%
\end{array}
\right.  \label{Heis}%
\end{equation}
\noindent This is the NC version of the Heisenberg algebra. In
this section we will develop an abstract representation theory of
this algebra. To this end we need two basic postulates:

\begin{enumerate}
\item There exists a base $\left|  x,p_{y}\right\rangle $ that
diagonalizes simultaneously both the position operator $\hat{x}$
and the momentum $\hat{p}_{y}$, whose domain of eigenvalues
coincides with all possible values of the coordinates $x$ and
$p_{y}$, which parameterize an arbitrary connected and
differentiable manifold $M$.
\begin{equation}%
\begin{array}
[c]{ccc}%
\hat{x}\left|  x,p_{y}\right\rangle =x\left|  x,p_{y}\right\rangle
, & \hat {p}_{y}\left|  x,p_{y}\right\rangle =p_{y}\left|
x,p_{y}\right\rangle , \\& x,\text{ }p_{y}\in M.
\end{array}
\end{equation}

\item There exists an internal product $\left\langle
\phi|\psi\right\rangle $, positive definite and Hermitian for
which the operators $\hat{x}$, $\hat{y}$, $\hat{p}_{x}$ and
$\hat{p}_{y}$ are self-adjoint.
\end{enumerate}
\bigskip

\noindent With the rules at hand, the application of the postulates
in the evaluation of the matrix elements $\left\langle x,p_{y}\left|
\hat{x}\right| x^{\prime},p_{y}^{\prime }\right\rangle $ and
$\left\langle x,p_{y}\left| \hat{p}_{y}\right| x^{\prime
},p_{y}^{\prime}\right\rangle $ imply:
\begin{align}
(x-x^{\prime})\left\langle
x,p_{y}|x^{\prime},p_{y}^{\prime}\right\rangle  &
=0,\\
(p_{y}-p_{y}^{\prime})\left\langle
x,p_{y}|x^{\prime},p_{y}^{\prime }\right\rangle  &  =0.\nonumber
\end{align}
\noindent The general solution to these equations is:
\begin{equation} \label{norm1}
\left\langle x,p_{y}|x^{\prime},p_{y}^{\prime}\right\rangle =\frac
{\delta(x-x^{\prime})\delta(p_{y}-p_{y}^{\prime})}{\sqrt{g(x,p_{y})}},
\end{equation}
\noindent where $g(x,p_{y})$ is a positive definite, arbitrary
function which is a \textsl{a priori} related to the normalization
of the eigenbasis. As a  consequence of this result the spectral
decomposition of the identity operator in the  $\left|
x,p_{y}\right\rangle $ base is:
\begin{equation}
\hat{1}=\int_{M}dxdp_{y}\sqrt{g(x,p_{y})}\left|
x,p_{y}\right\rangle \left\langle x,p_{y}\right|  .
\end{equation}
\noindent this completeness relation allow us to construct the
wave function:
\begin{equation}
\psi(x,p_{y})=\left\langle x,p_{y}|\psi\right\rangle , \label{fon1}%
\end{equation}
\noindent so that for any state $\left|  x,p_{y}\right\rangle $
belonging to the representation of the algebra  (\ref{Heis}),
\begin{align}
\left|  \psi\right\rangle  &
=\int_{M}dxdp_{y}\sqrt{g(x,p_{y})}\psi
(x,p_{y})\left|  x,p_{y}\right\rangle ,\\
\left\langle \psi\right|   &
=\int_{M}dxdp_{y}\sqrt{g(x,p_{y})}\psi^{\ast
}(x,p_{y})\left\langle x,p_{y}\right|  .\nonumber
\end{align}
\noindent In particular, the internal product of two arbitrary
states $\left| \psi\right\rangle $ and $\left|  \phi\right\rangle
$, is expressed in terms of the wave functions $\psi(x,p_{y})$ and
$\phi(x,p_{y})$:
\begin{equation}
\left\langle \psi|\phi\right\rangle =\int_{M}dxdp_{y}\sqrt{g(x,p_{y})}%
\psi^{\ast}(x,p_{y})\phi(x,p_{y}).
\end{equation}

\noindent Evaluation of the matrix elements of the commutators
$[\hat {x},\hat{y}]$ and $[\hat{y},\hat{p}_{y}]$, using the
postulates and Eq.(\ref{Heis}), lead to the relations:

\begin{align}
\left\langle x,p_{y}\left|  [\hat{x},\hat{y}]\right|  x^{\prime},p_{y}%
^{\prime}\right\rangle  &
=i\hbar\theta\frac{\delta(x-x^{\prime})\delta
(p_{y}-p_{y}^{\prime})}{\sqrt{g(x,p_{y})}}\\
&  =(x-x^{\prime})\left\langle x,p_{y}\left|  \hat{y}\right|  x^{\prime}%
,p_{y}^{\prime}\right\rangle ,\nonumber
\end{align}%
\begin{align}
\left\langle x,p_{y}\left|  [\hat{y},\hat{p}_{y}]\right|  x^{\prime}%
,p_{y}^{\prime}\right\rangle  &
=i\hbar\frac{\delta(x-x^{\prime})\delta
(p_{y}-p_{y}^{\prime})}{\sqrt{g(x,p_{y})}}\\
&  =-(p_{y}-p_{y}^{\prime})\left\langle x,p_{y}\left|
\hat{y}\right| x^{\prime},p_{y}^{\prime}\right\rangle .\nonumber
\end{align}
\noindent Thus, matrix element of the $\hat{y}$ operator can be
parameterized as follows:
\begin{align}
&  \left\langle x,p_{y}\left|  \hat{y}\right|
x^{\prime},p_{y}^{\prime
}\right\rangle \label{ye}\\
&  =\frac{i\hbar}{\sqrt{g(x,p_{y})}}\left(
-\theta\frac{\partial}{\partial x}+\frac{\partial}{\partial
p_{y}}\right)  \delta(x-x^{\prime})\delta
(p_{y}-p_{y}^{\prime})\nonumber\\
&  +\frac{\left[  A(x,p_{y})+iB(x,p_{y})\right]  }{\sqrt{g(x,p_{y})}}%
\delta(x-x^{\prime})\delta(p_{y}-p_{y}^{\prime}).\nonumber
\end{align}
\begin{widetext}
\noindent where $A(x,p_{y})$ and $B(x,p_{y})$ are two real,
arbitrary functions defined over $M$. Further constraints on $A$ and
$B$ can arise from the hermiticity of the $\hat{y}$ operator,
namely:
\begin{align} &
\frac{-i\hbar}{\sqrt{g(x,p_{y})}}\left(
-\theta\frac{\partial}{\partial x}+\frac{\partial}{\partial
p_{y}}\right)  \delta(x-x^{\prime})\delta (p_{y}-p_{y}^{\prime})
 +\frac{\left[  A(x,p_{y})-iB(x,p_{y})\right]  }{\sqrt{g(x,p_{y})}}%
\delta(x-x^{\prime})\delta(p_{y}-p_{y}^{\prime})\nonumber\\
&  =\frac{i\hbar}{\sqrt{g(x^{\prime},p_{y}^{\prime})}}\left(
-\theta
\frac{\partial}{\partial x^{\prime}}+\frac{\partial}{\partial p_{y}^{\prime}%
}\right)  \delta(x-x^{\prime})\delta(p_{y}-p_{y}^{\prime})\nonumber
  +\frac{\left[
A(x^{\prime},p_{y}^{\prime})+iB(x^{\prime},p_{y}^{\prime
})\right]  }{\sqrt{g(x^{\prime},p_{y}^{\prime})}}\delta(x-x^{\prime}%
)\delta(p_{y}-p_{y}^{\prime}).\nonumber
\end{align}
\noindent In order to solve this equation we assume the existence
of a continuous distribution $T(x^{\prime},p_{y}^{\prime})$ on
$M$, we multiply both sides of the last equation by $T(x^{\prime
},p_{y}^{\prime})$ and integrate over the domains of $x^{\prime}$
and $p_{y}^{\prime}$%
\begin{align}
&
\int_{M}T(x^{\prime},p_{y}^{\prime})dx^{\prime}dp_{y}^{\prime}\left\{
\frac{-i\hbar}{\sqrt{g(x,p_{y})}}\left(
-\theta\frac{\partial}{\partial x}+\frac{\partial}{\partial
p_{y}}\right)  \delta(x-x^{\prime})\delta
(p_{y}-p_{y}^{\prime})\right. \nonumber\\
&  \left.  +\frac{\left[  A(x,p_{y})-iB(x,p_{y})\right]  }{\sqrt{g(x,p_{y})}%
}\delta(x-x^{\prime})\delta(p_{y}-p_{y}^{\prime})\right\} \nonumber\\
&
=\int_{M}T(x^{\prime},p_{y}^{\prime})dx^{\prime}dp_{y}^{\prime}\left\{
\frac{i\hbar}{\sqrt{g(x^{\prime},p_{y}^{\prime})}}\left(  -\theta
\frac{\partial}{\partial x^{\prime}}+\frac{\partial}{\partial p_{y}^{\prime}%
}\right)  \delta(x-x^{\prime})\delta(p_{y}-p_{y}^{\prime})\right. \nonumber\\
&  \left.  +\frac{\left[  A(x^{\prime},p_{y}^{\prime})+iB(x^{\prime}%
,p_{y}^{\prime})\right]
}{\sqrt{g(x^{\prime},p_{y}^{\prime})}}\delta
(x-x^{\prime})\delta(p_{y}-p_{y}^{\prime})\right\}  ,
\end{align}
\noindent simplifying this expression and considering that
$T(x,p_{y})$ is an arbitrary function, $B(x,p_{y})$ turns out to
be:
\begin{equation}
B(x,p_{y})=\hbar\frac{\sqrt{g(x,p_{y})}}{2}\left(
-\theta\frac{\partial }{\partial x}+\frac{\partial}{\partial
p_{y}}\right)  \frac{1}{\sqrt {g(x,p_{y})}},
\end{equation}
\noindent therefore the matrix elements of $\hat{y}$ can be
expressed as:
\begin{align}
  \left\langle x,p_{y}\left|  \hat{y}\right|
x^{\prime},p_{y}^{\prime }\right\rangle \label{defA}
  =\frac{i\hbar}{g^{1/4}(x,p_{y})}\left(
-\theta\frac{\partial}{\partial
x}+\frac{\partial}{\partial p_{y}}\right)  \frac{\delta(x-x^{\prime}%
)\delta(p_{y}-p_{y}^{\prime})}{g^{1/4}(x,p_{y})}
  +\frac{A(x,p_{y})}{\sqrt{g(x,p_{y})}}\delta(x-x^{\prime})\delta(p_{y}%
-p_{y}^{\prime}).
\end{align}
\end{widetext}
\noindent The same approach can be applied to the operator
$\hat{p}_{x}$. To this end we consider the matrix elements of the
commutators $[\hat{x},\hat{p}_{x}]$ and
$[\hat{p}_{x},\hat{p}_{y}]$  to conclude that:
\begin{align}
\left\langle x,p_{y}\left|  \hat{p}_{x}\right|
x^{\prime},p_{y}^{\prime }\right\rangle  &
=\frac{-i\hbar}{g^{1/4}(x,p_{y})}\frac{\partial}{\partial
x}\left(  \frac{\delta(x-x^{\prime})\delta(p_{y}-p_{y}^{\prime})}%
{g^{1/4}(x,p_{y})}\right) \label{defC}\\
&  +\frac{C(x,p_{y})}{\sqrt{g(x,p_{y})}}\delta(x-x^{\prime})\delta(p_{y}%
-p_{y}^{\prime}),\nonumber
\end{align}
where $C(x,p_{y})$ is another arbitrary real function defined over
$M$.

\bigskip

\noindent An important consequence of the NC version of the
Heisenberg algebra (\ref{Heis}) arises from the evaluation of
\begin{equation}
\left\langle x,p_{y}\left|  [\hat{y},\hat{p}_{x}]\right|  x^{\prime}%
,p_{y}^{\prime}\right\rangle =0.
\end{equation}
The explicit calculation leads to the compatibility restriction
among the functions $A(x,p_{y})$ and $C(x,p_{y})$
\begin{equation}
\left(  -\theta\frac{\partial}{\partial x}+\frac{\partial}{\partial p_{y}%
}\right)  C(x,p_{y})+\frac{\partial}{\partial x}A(x,p_{y})=0. \label{relc}%
\end{equation}
If $\chi(x,p_{y})$\ is a scalar function defined over $M$, the
transformation
\begin{align}
C(x,p_{y})  &  \longrightarrow C(x,p_{y})-\frac{\partial}{\partial x}%
\chi(x,p_{y})\label{relctr}\\
A(x,p_{y})  &  \longrightarrow A(x,p_{y})+\left(
-\theta\frac{\partial }{\partial x}+\frac{\partial}{\partial
p_{y}}\right)  \chi(x,p_{y}),\nonumber
\end{align}
leaves the condition (\ref{relc}) invariant. \ This strongly
suggests that the functions $A(x,p_{y})$ and $C(x,p_{y})$ are
associated to the phase definition of the $\left|
x,p_{y}\right\rangle $ basis. Under a local U(1) gauge
transformation the states transform according to:
\begin{equation}
\left|  x,p_{y}\right\rangle
_{p}=e^{-\frac{i}{\hbar}\chi(\hat{x},\hat {p}_{y})}\left|
x,p_{y}\right\rangle ,
\end{equation}
the matrix elements in the transformed basis are related to the
original ones according to the following equations:
\begin{equation}
_{p}\left\langle x,p_{y}\left|  \hat{y}\right|
x^{\prime},p_{y}^{\prime }\right\rangle _{p}=\left\langle
x,p_{y}\left|  e^{\frac{i}{\hbar}\chi
(\hat{x},\hat{p}_{y})}\hat{y}e^{-\frac{i}{\hbar}\chi(\hat{x},\hat{p}_{y}%
)}\right|  x^{\prime},p_{y}^{\prime}\right\rangle \nonumber,
\end{equation}%
\begin{equation}
_{p}\left\langle x,p_{y}\left|  \hat{p}_{x}\right|  x^{\prime},p_{y}%
^{\prime}\right\rangle _{p}=\left\langle x,p_{y}\left|
e^{\frac{i}{\hbar
}\chi(\hat{x},\hat{p}_{y})}\hat{p}_{x}e^{-\frac{i}{\hbar}\chi(\hat{x},\hat
{p}_{y})}\right|  x^{\prime},p_{y}^{\prime}\right\rangle
\nonumber.
\end{equation}
The explicit evaluation of these relations yields
\begin{align}
&  _{p}\left\langle x,p_{y}\left|  \hat{y}\right|
x^{\prime},p_{y}^{\prime
}\right\rangle _{p}=\nonumber\\
&  =\left\langle x,p_{y}\left|  e^{\frac{i}{\hbar}\chi(\hat{x},\hat{p}_{y}%
)}\hat{y}e^{-\frac{i}{\hbar}\chi(\hat{x},\hat{p}_{y})}\right|
x^{\prime
},p_{y}^{\prime}\right\rangle \nonumber\\
&  =\frac{i\hbar}{g^{1/4}(x,p_{y})}\left(
-\theta\frac{\partial}{\partial
x}+\frac{\partial}{\partial p_{y}}\right)  \frac{\delta(x-x^{\prime}%
)\delta(p_{y}-p_{y}^{\prime})}{g^{1/4}(x,p_{y})}\nonumber\\
&  +\frac{\left[  A(x,p_{y})+\left(  -\theta\frac{\partial}{\partial x}%
+\frac{\partial}{\partial p_{y}}\right)  \chi(x,p_{y})\right]  }%
{\sqrt{g(x,p_{y})}}\delta(x-x^{\prime})\delta(p_{y}-p_{y}^{\prime}).\nonumber
\end{align}
and
\begin{align}
&  _{p}\left\langle x,p_{y}\left|  \hat{p}_{x}\right|  x^{\prime}%
,p_{y}^{\prime}\right\rangle _{p}\nonumber\\
&  =\left\langle x,p_{y}\left|  e^{\frac{i}{\hbar}\chi(\hat{x},\hat{p}_{y}%
)}\hat{p}_{x}e^{-\frac{i}{\hbar}\chi(\hat{x},\hat{p}_{y})}\right|
x^{\prime
},p_{y}^{\prime}\right\rangle \nonumber\\
&  =\frac{-i\hbar}{g^{1/4}(x,p_{y})}\frac{\partial}{\partial
x}\left(
\frac{\delta(x-x^{\prime})\delta(p_{y}-p_{y}^{\prime})}{g^{1/4}(x,p_{y}%
)}\right) \nonumber\\
&  +\frac{\left[  C(x,p_{y})-\frac{\partial}{\partial
x}\chi(x,p_{y})\right]
}{\sqrt{g(x,p_{y})}}\delta(x-x^{\prime})\delta(p_{y}-p_{y}^{\prime}),\nonumber
\end{align}
Thus, under phase transformations of the states, the functions
$A(x,p_{y})$ and $C(x,p_{y})$ behave according to (\ref{relctr}).
Therefore, the configuration space representations of the
Heisenberg algebra over the manifold $M$ are characterized, on one
hand, by the function $g(x,p_{y})$, and on the other  by a flat
U(1) bundle defined by the fields $A(x,p_{y})$ and $C(x,p_{y})$.
However, since arbitrary local gauge transformations within the
U(1) bundle correspond to arbitrary local phase redefinitions of
the states $\left|x,p_{y}\right\rangle $, and thus relate
representations of the Heisenberg algebra which are unitarily
equivalent, it is clear that all {\em inequivalent\/}
representations of the Heisenberg algebra over a manifold $M$ are
classified in terms of the topologically distinct flat U(1)
bundles over that manifold, {\em i.e.\/} the equivalence classes
under {\em local\/} gauge transformations of U(1) gauge fields of
vanishing field strength over $M$ \cite{gov}.
\bigskip

\noindent In the case of a simply connected manifold, every
holonomy is contractible to the identity.  Then, the gauge freedom
of $\left|  x,p_{y}\right\rangle $ can be used to remove both the
$A(x,p_{y})$ and $C(x,p_{y})$ fields through the adequate choice
of $\chi(x,p_{y})$.  Over a simply connected manifold, the NC
version of Heisenberg algebra admits only the representation in
which globally $A(x,p_{y})=0$ and $C(x,p_{y})=0$. When the base
manifold $M$ is not simply connected and therefore possess
topological obstructions that prevent some cycles to be
contracted, rendering non trivial holonomies around them, it is
not possible to completely remove both $A(x,p_{y})$ and
$C(x,p_{y})$.
\bigskip

\noindent Let us now turn our attention to the wave functions
$\psi (x,p_{y})=\left\langle x,p_{y}|\psi\right\rangle $.  Given
the parametrization of $\left\langle x,p_{y}\left| \hat{y}\right|
x^{\prime
},p_{y}^{\prime}\right\rangle $\ and $\left\langle x,p_{y}\left|  \hat{p}%
_{x}\right|  x^{\prime},p_{y}^{\prime}\right\rangle $, we can use
the spectral decomposition of the identity operator in order to obtain the
representation of $\hat{y}$ and $\hat{p}_{x}$ as differential operators
\begin{align}
&\left\langle x,p_{y}\left|  \hat{y}\right|  \psi\right\rangle   =
A(x,p_{y})\psi(x,p_{y})~~ +  \\
&\frac{i\hbar}{g^{1/4}(x,p_{y})}\left(
-\theta\frac{\partial}{\partial
x}+\frac{\partial}{\partial p_{y}}\right)  \left[  g^{1/4}(x,p_{y}%
)\psi(x,p_{y})\right] \nonumber ,
\end{align}%
\begin{align}
&\left\langle x,p_{y}\left|  \hat{p}_{x}\right|  \psi\right\rangle
= C(x,p_{y})\psi(x,p_{y})~~+ \\
& \frac{-i\hbar}{g^{1/4}(x,p_{y})}\frac{\partial}{\partial
x}\left[g^{1/4}(x,p_{y})\psi(x,p_{y})\right] \nonumber.
\end{align}
The last two  expressions are the general representation  of the
operators which takes into account: the NC, the measure functions
$g^{1/4}(x,p_{y})$ related to the inherent geometry of the base
manifold $M$ and to the overall normalization of the $\left|
x,p_{y}\right\rangle $ basis as well as the possibility of non
vanishing fields $A(x,p_{y})$ and $C(x,p_{y})$ which could arise
from the topological properties of $M$ and their possible
obstruction.

\bigskip

\noindent The properties  of the wave function $\psi(x,p_{y})$
required so that the operators $\hat{y}$ and $\hat{p}_{x}$\ are both
Hermitian and self-adjoint are
\begin{equation}
\int_{M}dxdp_{y}\left(  -\theta\frac{\partial}{\partial
x}+\frac{\partial
}{\partial p_{y}}\right)  \left[  \sqrt{g(x,p_{y})}|\left\langle x,p_{y}%
|\psi\right\rangle |^{2}\right]  =0
\end{equation}
and
\begin{equation}
\int_{M}dxdp_{y}\frac{\partial}{\partial x}\left[  \sqrt{g(x,p_{y}%
)}|\left\langle x,p_{y}|\psi\right\rangle |^{2}\right]  =0.
\end{equation}

\bigskip

\noindent Instead of using the basis $\left|  x,p_{y}\right\rangle
$, it is possible to work with sates $\left| y,p_{x}\right\rangle$
which diagonalizes simultaneously the position operator $\hat{y}$
and the momentum component $\hat{p}_{x}$  :
\begin{equation}%
\begin{array}
[c]{ccc}%
\hat{y}\left|  y,p_{x}\right\rangle =y\left|  y,p_{x}\right\rangle
, & \hat {p}_{x}\left|  y,p_{x}\right\rangle =p_{x}\left|
y,p_{x}\right\rangle , & y,p_{x}\in D(y,p_{x}).
\end{array}
\end{equation}
Where $D(y,p_{x})$ stands for the range of spectral values of $y$
and $p_{x}$. \
By analogy with the normalization of the $\left|  x,p_{y}%
\right\rangle $ eigenbasis (\ref{norm1}), the normalization of
$\left|  y,p_{x}%
\right\rangle $\ is parameterized according to the relation
\begin{equation}
\left\langle y,p_{x}|y^{\prime},p_{x}^{\prime}\right\rangle =\frac
{\delta(y-y^{\prime})\delta(p_{x}-p_{x}^{\prime})}{\sqrt{h(y,p_{x})}},
\end{equation}
where $h(y,p_{x})$\ is again an arbitrary positive definite
function defined over $D(y,p_{x})$. All the results are very
similar to the ones obtained in the $\left|  x,p_{y}\right\rangle$
basis, so instead of repeating the arguments we discuss the
quantities relevant to the change among those basis, {\it i.e.}
the wave functions $\left\langle x,p_{y}|y,p_{x}\right\rangle $. \
These wave functions are determined by the following set of
differential equations:
\begin{align}\label{set1}
&  \left[  -i\hbar\frac{\partial}{\partial x}+C(x,p_{y})\right]
\left[
g^{1/4}(x,p_{y})\left\langle x,p_{y}|y,p_{x}\right\rangle \right] \nonumber\\
&  =p_{x}\left[  g^{1/4}(x,p_{y})\left\langle
x,p_{y}|y,p_{x}\right\rangle \right]  ,
\end{align}%
\begin{align}
&  \left[  -i\hbar\left(  \theta\frac{\partial}{\partial
x}-\frac{\partial
}{\partial p_{y}}\right)  +A(x,p_{y})\right]  \left[  g^{1/4}(x,p_{y}%
)\left\langle x,p_{y}|y,p_{x}\right\rangle \right] \nonumber\\
&  =y\left[  g^{1/4}(x,p_{y})\left\langle
x,p_{y}|y,p_{x}\right\rangle \right]  .
\end{align}
Since the first order differential equations require only one an
integration constant, namely the wave function $\left\langle x^{0 }%
,p_{y}^{ 0  }|y,p_{x}\right\rangle $, associated to a specific
point on the manifold $M\ $\ of coordinates $(x^{ 0},p_{y}^{ 0
})$. \ Then, any other point of coordinates $(x,p_{y})$ can be
reached from $(x^{0 },p_{y}^{ 0 })$ through an oriented path
$P[(x^{0 },p_{y}^{0 })\longmapsto(x,p_{y})]$ running from $(x^{ 0
},p_{y}^{ 0 })$ to $(x,p_{y})$. \ The solution to (\ref{set1}) is
of the form
\begin{align}
&  g^{1/4}(x,p_{y})\left\langle x,p_{y}|y,p_{x}\right\rangle
=\left[ g^{1/4}(x^{0},p_{y}^{ 0 })\left\langle x^{0},p_{y}^{0
}|y,p_{x}\right\rangle
\right]  \times\nonumber\\
&  \times\Omega\left[  P[(x^{0 },p_{y}^{\left( 0\right)
})\longmapsto(x,p_{y})]\right]  \times\nonumber\\
&  \times e^{\frac{i}{\hbar}\left[  (x-x^{0 })p_{x}%
-(p_{y}-p_{y}^{ 0  })y+\theta(p_{y}-p_{y}^{\left( 0\right)
})p_{x}\right]  },
\end{align}
where $\Omega\left[  P[(x^{0 },p_{y}^{\left( 0\right)
})\longmapsto(x,p_{y})]\right]  $ represents an  ordered holonomy
along the path $P[(x^{\left( 0\right) },p_{y}^{\left( 0\right)
})\longmapsto(x,p_{y})]$ as shown by the following formula:
\begin{align}
&  \Omega\left[  P[(x^{0 },p_{y}^{ 0
})\longmapsto(x,p_{y})]\right]  =\label{Omeg1}\\
&  \exp\left\{  -\frac{i}{\hbar}\left[  \int_{P(x-x^{0 }%
)}dxC(x,p_{y})-\int_{P(p_{y}-p_{y}^{ 0  })}dp_{y}A(x,p_{y}%
)\right.  \right. \nonumber\\
&  \left.  \left.  +\theta\int_{P(p_{y}-p_{y}^{ 0  })}%
dp_{y}C(x,p_{y})\right]  \right\}  .\nonumber
\end{align}

\bigskip

\noindent The normalization condition of the wave function
$\left\langle y,p_{x}|y^{\prime},p_{x}^{\prime}\right\rangle $\
also requires that
\begin{equation}
\left|  g^{1/4}(x^{0 },p_{y}^{ 0 })\left\langle x^{0 },p_{y}^{ 0
}|y,p_{x}\right\rangle \right| ^{2}=\frac{1}{\left(
2\pi\hbar\right)  ^{2}\sqrt{h(y,p_{x})}},
\end{equation}
so that necessarily
\begin{equation}
g^{1/4}(x^{0 },p_{y}^{ 0 })\left\langle x^{\left( 0\right)
},p_{y}^{ 0 }|y,p_{x}\right\rangle
=\frac{e^{i\varphi(x^{0 },p_{y}^{ 0  },y,p_{x}%
)}}{(2\pi\hbar)h^{1/4}(y,p_{x})}.
\end{equation}
where $\varphi(x^{0 },p_{y}^{ 0 },y,p_{x})$\ is a specific real
function independent of $x$ and $p_{x}$. Then, the wave functions
$\left\langle x,p_{y}|y,p_{x}\right\rangle $\ are given by
\begin{align}
\left\langle x,p_{y}|y,p_{x}\right\rangle  & =\frac{e^{i\varphi(x^{0
},p_{y}^{ 0 },y,p_{x})}\Omega\left[ P[(x^{
0  },p_{y}^{ 0  })\longmapsto(x,p_{y})]\right]  }%
{(2\pi\hbar)g^{1/4}(x,p_{y})h^{1/4}(y,p_{x})}\times\nonumber\\
&  \times e^{\frac{i}{\hbar}\left[  (x-x^{0 })p_{x}%
-(p_{y}-p_{y}^{ 0  })y+\theta(p_{y}-p_{y}^{\left( 0\right)
})p_{x}\right]  }.
\end{align}
\noindent The specific choice of $\varphi(x^{\left( 0\right)
},p_{y}^{ 0 },y,p_{x})$, such that
\begin{equation}
e^{i\varphi(x^{0 },p_{y}^{ 0  },y,p_{x}%
)}e^{-\frac{i}{\hbar}\left[  x^{0) }p_{x}+p_{y}^{\left( 0\right)
}y-\theta p_{y}^{ 0 }p_{x}\right]  }=1
\end{equation}
\noindent simplifies the  wave function representation of NC
Heisenberg algebra:
\begin{equation}
\left\langle x,p_{y}|y,p_{x}\right\rangle =\frac{\Omega\left[ P[(x^{
0  },p_{y}^{ 0 })\longmapsto(x,p_{y})]\right]
e^{\frac{i}{\hbar}\left[  xp_{x}-p_{y}y+\theta p_{y}p_{x}\right]  }}%
{(2\pi\hbar)g^{1/4}(x,p_{y})h^{1/4}(y,p_{x})}. \label{ker1}%
\end{equation}
\noindent  This is the NC generalization of the customary wave
function ($\theta = 0$) that arises in conventional QM, and that
coincides with the Fourier Transform kernel.

\bigskip

\noindent Another admissible basis compatible with the commutation
relations (\ref{Heis}) is $\left| p\right\rangle =\left|
p_{x},p_{y}\right\rangle $, which diagonalizes simultaneously both
components of the momentum operator:
\begin{equation}%
\begin{array}
[c]{ccc}%
\hat{p}_{x}\left|  p\right\rangle =p_{x}\left|  p\right\rangle , & \hat{p}%
_{y}\left|  p\right\rangle =p_{y}\left|  p\right\rangle \text{.} &
p_{x}\text{, }p_{y}\in D(\text{\ }p).
\end{array}
\end{equation}
Where $D(p)$ is the range of spectral eigenvalues of $p_{x}$\ and
$p_{y}$. \ The normalization of$\left| p\right\rangle $ can be
parameterized according to
\begin{equation}
\left\langle p|p^{\prime}\right\rangle
=\frac{\delta(p_{x}-p_{x}^{\prime
})\delta(p_{y}-p_{y}^{\prime})}{\sqrt{\gamma(p_{x},p_{y})}},
\end{equation}
with $\gamma(p_{x},p_{y})$\ as a new arbitrary positive definite
function, defined on the $D(p)$ domain. \ This choice implies that
the spectral decomposition of the unit operator, in terms of the
momentum eigenbasis is of the form
\begin{equation}
\hat{1}=\int_{D(p)}dp_{x}dp_{y}\sqrt{\gamma(p_{x},p_{y})}\left|
p\right\rangle \left\langle p\right|  .
\end{equation}

\bigskip

\noindent A procedure similar to the one used in the  $\left|  x,p_{y}%
\right\rangle $, \ leads to the following matrix elements:
\begin{align}
\left\langle p\left|  \hat{x}\right|  p^{\prime}\right\rangle  &
=\frac{i\hbar}{\gamma^{1/4}(p_{x},p_{y})}\left(
\frac{\partial}{\partial
p_{x}}\right)  \frac{\delta(p_{x}-p_{x}^{\prime})\delta(p_{y}-p_{y}^{\prime}%
)}{\gamma^{1/4}(p_{x},p_{y})}\nonumber\\
&  +G_{x}(p_{x},p_{y})\frac{\delta(p_{x}-p_{x}^{\prime})\delta(p_{y}%
-p_{y}^{\prime})}{\sqrt{\gamma(p_{x},p_{y})}},\\
\left\langle p\left|  \hat{y}\right|  p^{\prime}\right\rangle  &
=\frac{i\hbar}{\gamma^{1/4}(p_{x},p_{y})}\left(
\frac{\partial}{\partial
p_{y}}\right)  \frac{\delta(p_{x}-p_{x}^{\prime})\delta(p_{y}-p_{y}^{\prime}%
)}{\gamma^{1/4}(p_{x},p_{y})}\nonumber\\
&  +G_{y}(p_{x},p_{y})\frac{\delta(p_{x}-p_{x}^{\prime})\delta(p_{y}%
-p_{y}^{\prime})}{\sqrt{\gamma(p_{x},p_{y})}}.
\end{align}
In these equations, $G_{x}(p_{x},p_{y})$ and $G_{y}(p_{x},p_{y})$
are the components of a vector field defined on $D(p)$, which by
virtue of the calculation of the matrix elements of the relation
$[x,y]=i\hbar\theta$, satisfy the compatibility relation
\begin{equation}
\frac{\partial G_{y}(p)}{\partial p_{x}}-\frac{\partial
G_{x}(p)}{\partial
p_{y}}=\theta. \label{ber}%
\end{equation}
This equation is invariant under the following transformation:
\begin{equation}
G_{i}(p)\longrightarrow G_{i}(p)+\frac{\partial\xi(p)}{\partial
p_{i}},
\end{equation}
with $\xi(p)$\ as a scalar local function defined on $D(p)$. \ It
is important to notice that, in contradistinction  to the
functions $A(x,p_{y})$ and $C(x,p_{y})$
Eqs.(\ref{defA},\ref{defC}), in the present case $G_{x}(p)=0$ and
$G_{y}(p)=0$ cannot be simultaneously satisfied. \ For this
reason, it is convenient to use a slight different
parametrization:
\begin{align}
\left\langle p\left|  \hat{x}\right|  p^{\prime}\right\rangle  &
=\frac{i\hbar}{\gamma^{1/4}(p_{x},p_{y})}\left(
\frac{\partial}{\partial
p_{x}}\right)  \frac{\delta(p_{x}-p_{x}^{\prime})\delta(p_{y}-p_{y}^{\prime}%
)}{\gamma^{1/4}(p_{x},p_{y})}\nonumber\\
&  +\left(  -\frac{\theta}{2}p_{y}+F_{x}(p_{x},p_{y})\right)
\frac
{\delta(p_{x}-p_{x}^{\prime})\delta(p_{y}-p_{y}^{\prime})}{\sqrt{\gamma
(p_{x},p_{y})}},\\
\left\langle p\left|  \hat{y}\right|  p^{\prime}\right\rangle  &
=\frac{i\hbar}{\gamma^{1/4}(p_{x},p_{y})}\left(
\frac{\partial}{\partial
p_{y}}\right)  \frac{\delta(p_{x}-p_{x}^{\prime})\delta(p_{y}-p_{y}^{\prime}%
)}{\gamma^{1/4}(p_{x},p_{y})}\nonumber\\
&  +\left(  \frac{\theta}{2}p_{x}+F_{y}(p_{x},p_{y})\right)  \frac
{\delta(p_{x}-p_{x}^{\prime})\delta(p_{y}-p_{y}^{\prime})}{\sqrt{\gamma
(p_{x},p_{y})}}.
\end{align}
With this convention, $F_{x}(p_{x},p_{y})$ and
$F_{y}(p_{x},p_{y})$ are also the components of a vector field
defined on $D(p)$ that satisfy the condition
\begin{equation}
\frac{\partial F_{y}(p)}{\partial p_{x}}-\frac{\partial
F_{x}(p)}{\partial p_{y}}=0,
\end{equation}
which is invariant under the transformations:
\begin{equation}
F_{i}(p)\longrightarrow F_{i}(p)+\frac{\partial\xi(p)}{\partial
p_{i}}.
\end{equation}

\bigskip

\noindent The $\hat p$ \  eigenfunction representation of the
operators $\hat{x}$\ and $\hat{y}$\ are:
\begin{align}
\left\langle p\left|  \hat{x}\right|  \psi\right\rangle  &
=\frac{i\hbar }{\gamma^{1/4}(p)}\left(  \frac{\partial}{\partial
p_{x}}\right)  \gamma
^{1/4}(p)\psi(p)\\
&  +\left(  -\frac{\theta}{2}p_{y}+F_{x}(p_{x},p_{y})\right)  \psi
(p),\nonumber
\end{align}%
\begin{align}
\left\langle p\left|  \hat{y}\right|  \psi\right\rangle  &
=\frac{i\hbar }{\gamma^{1/4}(p)}\left(  \frac{\partial}{\partial
p_{y}}\right)  \gamma
^{1/4}(p)\psi(p)\\
&  +\left(  \frac{\theta}{2}p_{x}+F_{y}(p_{x},p_{y})\right)
\psi(p).\nonumber
\end{align}
\bigskip

\noindent The wave function $\left\langle
x,p_{y}|p^{\prime}\right\rangle $\ can be determined through the
following set of differential equations
\begin{equation}
(p_{y}-p_{y}^{\prime})\left\langle x,p_{y}\right|
\widehat{p}_{x}\left| p^{\prime}\right\rangle =0,
\end{equation}%
\begin{align}
&  \left[  -i\hbar\frac{\partial}{\partial x}+C(x,p_{y})\right]
\left( g^{1/4}(x,p_{y})\left\langle
x,p_{y}|p^{\prime}\right\rangle \right)
\nonumber\\
&  =p_{x}^{\prime}g^{1/4}(x,p_{y})\left\langle x,p_{y}|p^{\prime
}\right\rangle
\end{align}
\begin{align}
&  \left[  -i\hbar\frac{\partial}{\partial p_{x}^{\prime}}-\frac{\theta}%
{2}p_{y}^{\prime}+F_{x}(p^{\prime})\right]  \left(
\gamma^{1/4}(p^{\prime
})\left\langle x,p_{y}|p^{\prime}\right\rangle \right) \nonumber\\
&  =x\gamma^{1/4}(p^{\prime})\left\langle
x,p_{y}|p^{\prime}\right\rangle .
\end{align}
The normalized solution to these equations is
\begin{align}
\left\langle x,p_{y}|p^{\prime}\right\rangle  &  =\frac{\delta(p_{y}%
-p_{y}^{\prime})e^{\frac{i}{\hbar}\left[  xp_{x}^{\prime}+\frac{\theta}%
{2}p_{y}^{\prime}p_{x}^{\prime}\right]  }e^{i\phi(x^{0 }%
,p_{y}^{ 0  },p_{x}^{\prime\left(  0\right)  },p_{x}%
^{\prime\left(  0\right)
})}}{\sqrt{2\pi\hbar}g^{1/4}(x,p_{y})\gamma
^{1/4}(p^{\prime})}\times\label{ker3}\\
&  \times\Xi\left[  P[(p_{x}^{\prime\left(  0\right)
},p_{y}^{\prime\left( 0\right)
})\longmapsto(p_{x}^{\prime},p_{y}^{\prime})]\right]  \times
\nonumber\\
&  \times\Omega\left[  P[(x^{0 },p_{y}^{\left( 0\right)
})\longmapsto(x,p_{y})]\right]  .\nonumber
\end{align}
where  $\Omega$ and $\Xi$ are holonomies along the ordered paths
connecting the fixed points $(x^{0) },p_{y}^{\left( 0\right) })$
and $(p_{x}^{\prime\left(  0\right) },p_{y}^{\prime\left( 0\right)
})$ to $(x,p_{y})$ and $(p_{x}^{\prime},p_{y}^{\prime})$,
respectively. \ $\Omega$ was defined before in (\ref{Omeg1}), and
$\Xi$ is given by:
\begin{align}
&  \Xi\left[  P[(p_{x}^{\prime\left(  0\right)
},p_{y}^{\prime\left(
0\right)  })\longmapsto(p_{x}^{\prime},p_{y}^{\prime})]\right]  =\\
&  e^{-\frac{i}{\hbar}\left[
\int_{P(p_{x}^{\prime}-p_{x}^{\prime\left(
0\right)  })}dp_{x}^{\prime}F_{x}(p^{\prime})+\int_{P(p_{y}^{\prime}%
-p_{y}^{\prime\left(  0\right)
})}dp_{y}^{\prime}F_{y}(p^{\prime})\right] },\nonumber
\end{align}
with $\phi(x^{0 },p_{y}^{ 0  },p_{x}%
^{\prime\left(  0\right)  },p_{x}^{\prime\left(  0\right)  })$\ as
a constant phase. \ Again, in the case of a simply connected base
manifold, the gauge freedom of the eigenstates $\left|
p\right\rangle $ can be used to remove completely the vector field
$F_{i}(p)$ through the correct choice of the gauge transformation.

\bigskip

\noindent The wave function $\left\langle
x,p_{y}|p^{\prime}\right\rangle $ can be constructed from
$\left\langle x,p_{y}|y,p_{x}\right\rangle $, given by
(\ref{ker1}), and from the spectral decomposition of the unity
operator:
\begin{align}
& \left\langle y,p_{x}|p^{\prime}\right\rangle   =\int_{M}dxdp_{y}%
\sqrt{g(x,p_{y})}\left\langle y,p_{x}|x,p_{y}\right\rangle
\left\langle
x,p_{y}|p^{\prime}\right\rangle \label{ker2}\\
&  =\frac{\delta(p_{x}-p_{x}^{\prime})e^{\frac{i}{\hbar}\left[  yp_{y}%
^{\prime}-\frac{\theta}{2}p_{y}^{\prime}p_{x}^{\prime}\right]  }%
e^{i\Phi(y,p_{x},p_{x}^{\prime\left(  0\right)
},p_{x}^{\prime\left(
0\right)  })}}{\sqrt{2\pi\hbar}h^{1/4}(x,p_{y})\gamma^{1/4}(p^{\prime})}%
\times\nonumber\\
&  \times\Xi\left[  P[(p_{x}^{\prime\left(  0\right)
},p_{y}^{\prime\left( 0\right)
})\longmapsto(p_{x}^{\prime},p_{y}^{\prime})]\right]  ,\nonumber
\end{align}
where $\Phi(y,p_{x},p_{x}^{\prime\left(  0\right)
},p_{x}^{\prime\left( 0\right)  })$ is an arbitrary real function.

\bigskip

\noindent To conclude this section we quote the change of basis
among the different wave functions $\Psi(x,p_{y}%
)=\left\langle x,p_{y}|\Psi\right\rangle $,
$\Psi(y,p_{x})=\left\langle y,p_{x}|\Psi\right\rangle $ and
$\Psi(p)=\left\langle p|\Psi\right\rangle $:\
\begin{align}
\Psi(x,p_{y})  &
=\int_{D(y,p_{x})}dydp_{x}\sqrt{h(y,p_{x})}\left\langle
x,p_{y}|y,p_{x}\right\rangle \Psi(y,p_{x}),\nonumber\label{transf}\\
\Psi(x,p_{y})  &  =\int_{D(p_{x},p_{y})}dp_{x}^{\prime}dp_{y}^{\prime}%
\sqrt{\gamma(p_{x}^{\prime},p_{y}^{\prime})}\left\langle
x,p_{y}|p^{\prime
}\right\rangle \Psi(p^{\prime}),\nonumber\\
\Psi(y,p_{x})  &  =\int_{M}dxdp_{y}\sqrt{g(x,p_{y})}\left\langle
y,p_{x}|x,p_{y}\right\rangle \Psi(x,p_{y}),\nonumber\\
\Psi(y,p_{x})  &  =\int_{D(p_{x},p_{y})}dp_{x}^{\prime}dp_{y}^{\prime}%
\sqrt{\gamma(p_{x}^{\prime},p_{y}^{\prime})}\left\langle
y,p_{x}|p^{\prime
}\right\rangle \Psi(p^{\prime}),\nonumber\\
\Psi(p^{\prime})  &
=\int_{D(y,p_{x})}dydp_{x}\sqrt{h(y,p_{x})}\left\langle
p^{\prime}|y,p_{x}\right\rangle \Psi(y,p_{x})\\
\Psi(p^{\prime})  &
=\int_{M}dxdp_{y}\sqrt{g(x,p_{y})}\left\langle p^{\prime
}|x,p_{y}\right\rangle \Psi(x,p_{y}),\nonumber
\end{align}
with $\left\langle x,p_{y}|y,p_{x}\right\rangle $, $\left\langle x,p_{y}%
|p^{\prime}\right\rangle $\ and $\left\langle
y,p_{x}|p^{\prime}\right\rangle $ given by equations (\ref{ker1}),
(\ref{ker3}) and (\ref{ker2}), respectively.

\bigskip

\section{Path Integral in Phase Space}

\bigskip

\noindent In this section, we show the relation between the
canonical quantization formalism discussed in the previous part
and the path integral representation of quantum amplitudes. We
follow the conventional approach as well as a previous work
devoted to the NC case \cite{aci}
\bigskip

\noindent Given a system with initial configuration $i$, the
probability associated to the evolution of this system towards a
final configuration $f$ is
\begin{equation}
P_{f\leftarrow i}=\left|  \left\langle f\right|
\hat{U}(t_{f},t_{i})\left| i\right\rangle \right|  ^{2}.
\end{equation}
If we choose the eigenbasis $\left|  x,p_{y}\right\rangle $ to
label initial and final states, the transition amplitude can be
written as
\begin{equation}
K(x_{f},p_{yf},t_{f};x_{i},p_{yi},t_{i})=\left\langle
x_{f},p_{yf}\right| e^{-\frac{i}{\hbar}(t_{f}-t_{i})\hat{H}}\left|
x_{i},p_{yi}\right\rangle .
\end{equation}

\bigskip

\noindent The argument is based on the factorization of the
temporal evolution operator in the form:
\[
\hat{U}(t_{f},t_{i})=\left\{  e^{\left[  -\frac{i}{\hbar}\frac{(t_{f}-t_{i}%
)}{N}\hat{H}\right]  }\right\}  ^{N}.
\]
Inserting two spectral decompositions of the unity operator (in
the $\left| x,p_{y}\right\rangle $ and $\left|
y,p_{x}\right\rangle $ basis) between each of the $N$ factors and
making use of the wave functions $\left\langle
x_{j},p_{yj}|y_{k},p_{xk}\right\rangle$ Eq.(\ref{ker1}), the
Kernel can be written as:

\begin{align}
&  K(x_{f},p_{yf},t_{f};x_{i},p_{yi},t_{i})\label{PI1}\\
&  =\lim_{N\rightarrow\infty}\left\langle x_{f},p_{yf}\right| \left(
1-\frac{i}{\hbar}\epsilon\hat{H}\right)  ^{N}\left|
x_{i},p_{yi}\right\rangle
\nonumber\\
&  =\lim_{N\rightarrow\infty}\prod_{j=1}^{N-1}\int_{M}dx_{j}dp_{yj}%
\sqrt{g_j}\prod_{k=0}^{N-1}\left[  \int_{D_k}%
dy_{k}dp_{xk}\times\right. \nonumber\\
&  \left.  \times\sqrt{h_k}\left\langle x_{k+1},p_{yk+1}%
|y_{k},p_{xk}\right\rangle \left\langle y_{k},p_{xk}\right|
1-\frac{i}{\hbar
}\epsilon\hat{H}\left|  x_{k},p_{yk}\right\rangle \right] \nonumber\\
&  =\lim_{N\rightarrow\infty}\prod_{j=1}^{N-1}\int_{M}dx_{j}dp_{yj}%
\sqrt{g_j}\prod_{k=0}^{N-1}\left[  \int_{D_k}%
dy_{k}dp_{xk}\times\right. \nonumber\\
&  \times\sqrt{h_k}\frac{\Omega\left[ P[(x_{k+1}^{
0  },p_{yk+1}^{ 0 })\longmapsto(x_{k+1},p_{yk+1}%
)]\right]  }{(2\pi\hbar)^{2}g^{1/4}(x_{k+1},p_{yk+1})h^{1/4}(y_{k},p_{xk}%
)}\times\nonumber\\
&  \times\frac{\Omega^{-1}\left[  P[(x_{k}^{ 0  },p_{yk}%
^{ 0 })\longmapsto(x_{k},p_{yk})]\right]  }{g^{1/4}%
(x_{k},p_{yk})h^{1/4}(y_{k},p_{xk})}\left( 1-\frac{i}{\hbar}\epsilon
h_{k}\right)\nonumber\\
&  \left.  \times e^{\frac{i}{\hbar}\left[  \left(
x_{k+1}-x_{k}\right)
p_{xk}-\left(  p_{yk+1}-p_{yk}\right)  y_{k}+\theta\left(  p_{yk+1}%
-p_{yk}\right)  p_{xk}\right]  }  \right]  ,\nonumber
\end{align}
where, to avoid lengthly expressions, we introduced the following
notation:
\begin{equation} g_j=g(x_{j},p_{yj}),~~ h_j=h(y_{j},p_{xj}), ~~D_k=D(y_{k},p_{xk}) \nonumber
\end{equation}
\begin{equation}%
\begin{array}
[c]{c}%
\begin{array}
[c]{cc}%
\epsilon=\frac{t_{f}-t_{i}}{N}, & h_{i}=\frac{\left\langle y_{i}%
,p_{xi}\right|  \hat{H}\left|  x_{i},p_{yi}\right\rangle
}{\left\langle y_{i},p_{xi}|x_{i},p_{yi}\right\rangle
}=h_{i}^{\ast},
\end{array}
\\
i=0,1,^{...},N-1
\end{array}
\end{equation}

\bigskip

\noindent Simplifying this expression, we arrive to
\begin{align}
&  \left\langle x_{f},p_{yf}\right|  \hat{U}(t_{f},t_{i})\left|  x_{i}%
,p_{yi}\right\rangle =\Omega^{-1}\left[P[(x_{i}^{0 },p_{yi}^{
0})\mapsto(x_{i},p_{yi})]\right]  \nonumber\\
&  \frac{\Omega\left[  P[(x_{f}^{ 0 },p_{yf}^{ 0
})\longmapsto(x_{f},p_{yf})]\right]  }{(2\pi\hbar)^{2}g^{1/4}(x_{f}%
,p_{yf})g^{1/4}(x_{i},p_{yi})}\times\nonumber\\
&  \times\lim_{N\rightarrow\infty}\int_{M}\prod_{j=1}^{N-1}dx_{j}dp_{yj}%
\int_{D(y_{j},p_{xj})}\prod_{j=0}^{N-1}dy_{j}dp_{xj}\times\nonumber\\
&  \times e^{\frac{i}{\hbar}\sum_{j=0}^{N-1}\epsilon\left[
\frac{\left( x_{j+1}-x_{j}\right)  }{\epsilon}p_{xj}-\frac{\left(
p_{yj+1}-p_{yj}\right)
}{\epsilon}y_{j}+\theta\frac{\left(  p_{yj+1}-p_{yj}\right)  }{\epsilon}%
p_{xj}-h_{j}\right]  }.\nonumber
\end{align}
As expected, factors
\begin{align}
&  \Omega\left[  P[(x_{k+1}^{\left(  0\right)  },p_{yk+1}^{\left(
0\right)
})\longmapsto(x_{k+1},p_{yk+1})]\right]  \times\nonumber\\
&  \times\Omega^{-1}\left[  P[(x_{k}^{\left(  0\right)
},p_{yk}^{\left( 0\right)  })\longmapsto(x_{k},p_{yk})]\right]
\end{align}
in (\ref{PI1}) cancel out among themselves, except for those
associated to $(x_{f},p_{yf})$  and  $(x_{i},p_{yi}).$

\bigskip

\noindent Finally, the kernel expressed as Functional Integrals
over phase space (up to irrelevant normalization factors) is:
\begin{align}
&  \left\langle x_{f},p_{yf}\right|  \hat{U}(t_{f},t_{i})\left|  x_{i}%
,p_{yi}\right\rangle \\
&  =\int_{x_{i};p_{yi}}^{x_{f}, p_{yf}}\mathcal{D}x\mathcal{D}y\mathcal{D}p_{x}\mathcal{D}%
p_{y}e^{\frac{i}{\hbar}\int_{t_{i}}^{t_{f}}dt\left[  \dot{x}p_{x\text{ }}%
-\dot{p}_{y}y+\theta\dot{p}_{y}p_{x\text{ }}-H\right]  },\nonumber
\end{align}
where the appropriate boundary conditions for the functional
integrals has been indicated using the notation $x(t_{i})=x_{i},
p_{y}(t_{i})=p_{yi}, {x(t_{f})=x_{f}, p_{y} (t_{f})=p_{yf}}$. \ In
this formal expression, the integration measure over phase space is
the Liouville measure. \ One can easily identify in the last
relation the classical action up to an irrelevant surface term
\begin{align}
&  \int_{t_{i}}^{t_{f}}dt\left[  \dot{x}p_{x\text{
}}-\dot{p}_{y}y+\theta
\dot{p}_{y}p_{x\text{ }}-H\right] \\
&  =\int_{t_{i}}^{t_{f}}dt\left[
\frac{1}{2}z^{\alpha}\omega_{\alpha\beta
}\dot{z}^{\beta}-H(z,t)+\frac{d\Lambda}{dt}\right] \nonumber\\
&  =S[z(t)]+\Lambda|_{t_{i}}^{t_{f}}.\nonumber
\end{align}
\noindent It should be clear that the path integral representation
can be expressed not only using the $\left| x ,p_{y}\right\rangle$
states but in any of the basis we analyzed.

\bigskip

\section{Wigner Function in Phase Space}

\bigskip

\noindent In this section we discuss the third independent, and
complete, description of QM, formally distinct to the conventional
operator approach in Hilbert space and to the Path integral
quantization procedure. \ This quantization framework is based on
the Wigner quasi-distribution  function \cite{Wigner1}. \ The main
feature of this formalism is the fact that interprets the
coordinates of phase space $z^{\alpha}$ not as operators but as
c-numbers.

\bigskip

\noindent Wigner's function can be built from the density matrix.
The density operator is defined as the weighted sum over all
possible projectors%
\begin{equation}
\hat{\rho}=\sum_{n}w_{n}\left|  \phi_{n}\right\rangle \left\langle
\phi _{n}\right|  ,
\end{equation}
the $\left|\phi_{n}\right\rangle $~form a complete set of
normalized states and
\begin{equation}
\sum_{n}w_{n}=1.\label{norma}
\end{equation}

\noindent The matrix elements of this operator, with respect to
the $\left| x,p_{y}\right\rangle $ basis, for instance, are
\begin{align}
\rho(x,p_{y};x^{\prime},p_{y}^{\prime})  &
=\sum_{n}w_{n}\left\langle x,p_{y}|\phi_{n}\right\rangle
\left\langle \phi_{n}|x^{\prime},p_{y}^{\prime
}\right\rangle \\
&
=\sum_{n}w_{n}\phi_{n}(x,p_{y})\phi_{n}^{\ast}(x^{\prime},p_{y}^{\prime
}).\nonumber
\end{align}
Due to the normalization of the states and to Eq.(\ref{norma}):
\begin{equation}
Tr(\hat{\rho})=\int\rho(x,p_{y};x,p_{y})dxdp_{y}=1.
\end{equation}
\qquad\qquad\

\bigskip

\noindent For a given operator $\hat{A}$, the ensemble expected
value \cite{LandSP} is defined as
\begin{align}
\overline{\left\langle \hat{A}\right\rangle }  &  =Tr(\hat{\rho}\hat{A}%
)\\
&  =\sum_{n}w_{n}\int\phi_{n}(x,p_{y})\left(  \hat{A}\phi_{n}^{\ast}%
(x,p_{y})\right)  dxdp_{y}.\nonumber
\end{align}

\noindent If $w_{j}=1$ and $w_{i\neq j}=0$ the system is in a pure
state; otherwise, the system is in a mixed state. \ The quantity
$Tr(\hat{\rho}^{2})=$ $\sum_{n}w_{n}^{2}\leq1$ (with
$Tr(\hat{\rho}^{2})=1$ only possible for pure states), is called
Purity.

\bigskip

The quasi-distribution Wigner function is defined through
Wigner-Weyl prescription, which assigns a c-number function
$A_{W}(z)$ to each operator  $\hat{A}$ in Hilbert space. \ For the
two dimensional under consideration, we have explicitly:
\begin{align}
A_{W}(z)  &  =\frac{1}{(2\pi\hbar)^{4}}\int
d^2\sigma d^2\tau \left\{  e^{\frac{i}{\hbar}\Phi(\sigma,\tau)}\times\right. \label{Weyl1}\\
&  \left.  \times Tr\left[
e^{-\frac{i}{\hbar}(\tau_{1}\hat{p}_{x}+\tau
_{2}\hat{p}_{y}+\hat{\sigma}_{1}x+\hat{\sigma}_{2}y)}\hat{A}\right]
\right\} .\nonumber
\end{align}
where we introduced the notation $d^2\sigma d^2\tau=
d\sigma_{1}d\sigma_{2}d\tau_{1} d\tau_{2}$ and $\Phi(\sigma,\tau)
= \tau_{1}p_{x}+\tau_{2}p_{y} +\sigma_{1}x+\sigma_{2}y$. The
Wigner function $W(z)$ is defined as:
\begin{align}
W(z)  &  =\frac{1}{(2\pi\hbar)^{4}}\int d^2\sigma d^2\tau \left\{
e^{\frac{i}{\hbar}\Phi(\sigma,\tau)}\times\right. \label{Wig1}\\
&  \left.  \times Tr\left[
e^{-\frac{i}{\hbar}(\tau_{1}\hat{p}_{x}+\tau
_{2}\hat{p}_{y}+\hat{\sigma}_{1}x+\hat{\sigma}_{2}y)}\hat{\rho}\right]
\right\} \nonumber\\
&  =\frac{1}{(2\pi\hbar)^{4}}\int d^2\sigma d^2\tau \left\{
e^{\frac{i}{\hbar}\Phi(\sigma,\tau)}\int_{M}dx^{\prime}dp_{y}^{\prime}
\times\right. \nonumber\\
&  \left.  \times\left\langle x^{\prime
},p_{y}^{\prime}\left|  e^{-\frac{i}{\hbar}(\tau_{1}\hat{p}_{x}+\tau_{2}%
\hat{p}_{y}+\sigma_{1}\hat{x}+\sigma_{2}\hat{y})}\hat{\rho}\right|
x^{\prime },p_{y}^{\prime}\right\rangle \right\}  .\nonumber
\end{align}
\bigskip

\noindent In the following we will restraint to a cartesian and
simply connected NC phase space, so that $g(x,p_{y})=1$,
$h(y,p_{y})=1$, $\gamma(p)=1$ and we can also remove the functions
$A(x,p_{y})$, $C(x,p_{y})$ and $F_{i}(p)$ by means of a local
gauge transformations.
\bigskip

\noindent Using the commutation relations Eq.(\ref{Heis}) and the
representation of the operators Eq.(\ref{defA},\ref{defC}) and
analogous equations not explicitly written,  it follows that the
operators $\hat{x},$ $\hat{y},$ $\hat{p}_{x}$ and $\hat{p}_{y}$,
are generators of translations in phase space:
\begin{align}
e^{-\frac{i}{\hbar}a\hat{p}_{x}}\left|  x,p_{y}\right\rangle  &
=\left|
x+a,p_{y}\right\rangle ,\label{expo}\\
e^{-\frac{i}{\hbar}b\hat{y}}\left|  x,p_{y}\right\rangle  &
=\left|  x+\theta
b,p_{y}-b\right\rangle ,\nonumber\\
e^{-\frac{i}{\hbar}c\hat{p}_{y}}\left|  y,p_{x}\right\rangle  &
=\left|
y+c,p_{x}\right\rangle ,\nonumber\\
e^{-\frac{i}{\hbar}d\hat{x}}\left|  y,p_{x}\right\rangle  &
=\left|  y-\theta
d,p_{x}-d\right\rangle ,\nonumber\\
e^{-\frac{i}{\hbar}f\hat{x}}\left|  p_{x},p_{y}\right\rangle  &
=\left|
p_{x}-f,p_{y}\right\rangle ,\nonumber\\
e^{-\frac{i}{\hbar}g\hat{y}}\left|  p_{x},p_{y}\right\rangle  &
=\left| p_{x},p_{y}-g\right\rangle ,\nonumber
\end{align}
where $a$, $b$, $c$, $d$, $f$ and $g$ are arbitrary constants.
Using the Baker-Campbell-Hausdorff formula
\begin{equation}
e^{\hat{A}+\hat{B}}=e^{\hat{A}}e^{\hat{B}}e^{-\frac{1}{2}\left[  \hat{A}%
,\hat{B}\right]  +\cdots},
\end{equation}
the Wigner function is written as:
\begin{align}
&  W(z)=\int \frac{d^2\sigma d^2\tau}{(2\pi\hbar)^{4}} \left\{
e^{\frac{i}{\hbar}(\Phi(\sigma,\tau)-
\tau_{1}\sigma_{1}/2-\sigma_{1}\sigma_{2}\theta/2+\tau
_{2}\sigma_{2}/2)}\right.  \times\nonumber\\
&  \times\int dx^{\prime}dp_{y}^{\prime}\left\langle
x^{\prime},p_{y}^{\prime
}\right|  e^{-\frac{i}{2\hbar}\tau_{1}\hat{p}_{x}}e^{-\frac{i}{\hbar}%
\sigma_{2}\hat{y}}e^{-\frac{i}{\hbar}\sigma_{1}\hat{x}}e^{-\frac{i}{\hbar}%
\tau_{2}\hat{p}_{y}}\times\nonumber\\
&  \left.  \times\hat{\rho}e^{-\frac{i}{2\hbar}\tau_{1}\hat{p}_{x}}%
e^{-\frac{i}{2\hbar}\sigma_{2}\hat{y}}\left|
x^{\prime},p_{y}^{\prime }\right\rangle \right\}  .\nonumber
\end{align}
\newline This can be still simplified using the operators as translation
generators:
\begin{align}
&  W(z)=\int \frac{ d\sigma_{2}d\tau_{1}}{(2\pi\hbar)^{2}} \left\{
e^{\frac
{i}{\hbar}(\tau_{1}p_{x}+\sigma_{2}y)}  \right.\nonumber \\
&  \left.  \left\langle x-\frac{\tau_{1}}{2}-\theta\frac{\sigma_{2}}{2}%
,p_{y}+\frac{\sigma_{2}}{2}\left|  \hat{\rho}\right|  x+\frac{\tau_{1}}%
{2}+\theta\frac{\sigma_{2}}{2},p_{y}-\frac{\sigma_{2}}{2}\right\rangle
\right\}  .\nonumber
\end{align}

\bigskip

\noindent The change of the integration variables
$\zeta=\tau_{1}/2+\theta\sigma_{2}/2$, $\eta=-\sigma_{2}/2$ $\ $
is useful to write Wigner function in a more compact way
\begin{align}
W(z)  &  =\frac{1}{\pi^{2}\hbar^{2}}\int d\zeta d\eta\left\{
e^{\frac
{2i}{\hbar}(\zeta p_{x}-\eta y+\theta\eta p_{x})}\times\right. \label{Wig2}\\
&  \left.  \times\left\langle x-\zeta,p_{y}-\eta\left|
\hat{\rho}\right| x+\zeta,p_{y}+\eta\right\rangle \right\}
.\nonumber
\end{align}
If the system is in a pure state, with wave function
$\Psi(x,p_{y};t)$, then Wigner function takes the form
\begin{align}
W_{\Psi}(z)  &  =\frac{1}{\pi^{2}\hbar^{2}}\int d\zeta
d\eta\left\{
e^{\frac{2i}{\hbar}(\zeta p_{x}-\eta y+\theta\eta p_{x})}\times\right. \\
&  \left.  \times\Psi\left(  x-\zeta,p_{y}-\eta;t\right)
\Psi^{\ast}\left( x+\zeta,p_{y}+\eta;t\right)  \right\}
.\nonumber
\end{align}

\bigskip

\noindent It is also possible to define Wigner function $W(z)$
starting from the $\left| y,p_{x}\right\rangle $ basis using
(\ref{Wig2}, \ref{ker1}) and the spectral decomposition of the unity
operator;
\begin{align}
W(z)  &  =\frac{1}{\pi^{2}\hbar^{2}}\int dudv\left\{
e^{-\frac{2i}{\hbar
}(xu-p_{y}v+\theta up_{y})}\times\right. \\
&  \left.  \times\left\langle y-v,p_{x}-u\left|  \hat{\rho}\right|
y+v,p_{x}+u\right\rangle \right\}  .\nonumber
\end{align}
Similarly, in terms of the $\left| p\right\rangle $ basis the
Wigner function takes the form:
\begin{align}
W(z)  &  =\frac{1}{\pi^{2}\hbar^{2}}\int dud\eta\left\{
e^{-\frac{2i}{\hbar }\left[
xu+y\eta-\frac{\theta}{2}(p_{x}\eta-p_{y}u)\right]  }\times\right.
\label{Wp}\\
&  \left.  \times\left\langle p_{x}-u,p_{y}-\eta\left|
\hat{\rho}\right| p_{x}+u,p_{y}+\eta\right\rangle \right\}
.\nonumber
\end{align}

\bigskip

\noindent In analogy to the commutative case ($\theta = 0$), the
main features of $W(z)$ are:

\begin{enumerate}
\item  Wigner function $W(z)$ is real
\begin{equation}
W(z)^{\ast}=W(z)\nonumber.
\end{equation}

\item  If integrated over $x$ and $p_{y}$, $W_{\Psi}(z)$ gives the
correct marginal probability distribution on $y$ and $p_{x}$:
\begin{equation}
\left|  \Psi(y,p_{x})\right|  ^{2}=\int W_{\Psi}(z)dxdp_{y}
\nonumber.
\end{equation}
Similarly, if integrated over $y$ and $p_{x}$, the Wigner function
reproduces the probability distribution on $x$ and $p_{y}$:
\begin{equation}
\left|  \Psi(x,p_{y})\right|  ^{2}=\int W_{\Psi}(z)dydp_{x}
\nonumber.
\end{equation}
Finally, in order to obtain the marginal probability distribution
on the momentum components, it is sufficient to integrate
$W_{\Psi}(z)$ over $p_{x}$ and $p_{y}$:
\begin{equation}
\left|  \Psi(p)\right|  ^{2}=\int W_{\Psi}(z)dp_{x}dp_{y}
\nonumber.
\end{equation}
It is important to remark that wave functions$\Psi(p)$,
$\Psi(y,p_{x})$ and $\Psi (x,p_{y})$ are related by means of the
transformations (\ref{transf}).

\item  A consequence of the previous feature of Wigner function,
it is evident that $W_{\Psi}(z)$ is normalized
\[
\int W_{\Psi}(z)dxdydp_{x}dp_{y}=1.
\]

\item  Starting from two different density operators
$\hat{\rho}_{1}$ and $\hat{\rho}_{2}$, it is possible to construct
two different Wigner functions $W_{1}(z)$ and $W_{2}(z)$. \ The
operation $Tr(\hat{\rho}_{1}\hat{\rho}_{2})$, in terms of
$W_{1}(z)$ and $W_{2}(z)$, is given by
\begin{equation}
Tr(\hat{\rho}_{1}\hat{\rho}_{2})=\left(  2\pi\hbar\right) ^{2}\int
W_{1}(z)W_{2}(z)dxdydp_{x}dp_{y} \nonumber.
\end{equation}
Thus, if $A_{W}(z)$ is a Wigner function associated to the
operator $\hat{A}$ (\ref{Weyl1}):
\begin{align}
A_{W}(z)  &  =\frac{1}{\pi^{2}\hbar^{2}}\int d\zeta d\eta\left\{
e^{\frac
{2i}{\hbar}(\zeta p_{x}-\eta y+\theta\eta p_{x})}\times\right.\nonumber \\
&  \left.  \times\left\langle x-\zeta,p_{y}-\eta\left|
\hat{A}\right| x+\zeta,p_{y}+\eta\right\rangle \right\} ,\nonumber
\end{align}
then, the ensemble mean value of $\hat{A}$ is
\[
\overline{\left\langle \hat{A}\right\rangle
}=Tr(\hat{\rho}\hat{A})=\left( 2\pi\hbar\right)  ^{2}\int
W(z)A_{W}(z)dxdydp_{x}dp_{y} \nonumber.
\]

\item  If a system is in the state $\left| \psi\right\rangle $,
and a measurement that determines that the new state of the system
is $\left| \phi\right\rangle$ , then the probability to obtain
this result from the measurement is $\left| \left\langle
\psi|\phi\right\rangle \right| ^{2}$. \ In terms of Wigner
functions, the transition probability can be written as
\begin{equation}
\left|  \left\langle \psi|\phi\right\rangle \right|  ^{2}=\left(
2\pi \hbar\right)  ^{2}\int
W_{\psi}(z)W_{\phi}(z)dxdydp_{x}dp_{y}\nonumber.
\end{equation}
This expression can be interpreted as the proof that Wigner
function cannot be positive definite over phase space.\ If $\psi$
and $\phi$ are orthogonal, the last integral must vanish, and
therefore, if $W_{\phi}(z)$ is not equal to zero in a specific
region of phase state, then it must take negative values in
another sector. \
\end{enumerate}

\bigskip

\noindent The time dependence on Wigner function follows from:
\begin{align}
&  i\hbar\frac{\partial}{\partial t}W(z,t) = \int \frac{i\hbar
~d\zeta d\eta}{\pi^{2}\hbar^{2}} \left\{ e^{\frac
{2i}{\hbar}(\zeta p_{x}-\eta y+\theta\eta p_{x})}\times\right. \nonumber\\
&  \times\left[  \frac{\partial\Psi\left(
x-\zeta,p_{y}-\eta;t\right) }{\partial t}\Psi^{\ast}\left(
x+\zeta,p_{y}+\eta;t\right)  \right.
\nonumber\\
&  \left.  \left.  +\Psi\left(  x-\zeta,p_{y}-\eta;t\right)  \frac
{\partial\Psi^{\ast}\left(  x+\zeta,p_{y}+\eta;t\right) }{\partial
t}\right] \right\}  .\nonumber
\end{align}

\noindent the time dependent Schr\"{o}dinger equation can be used in
this expression. \ In the ($x,p_{y}$) representation the equation
including a potential $V(\hat{x},\hat{y})$ reads:
\begin{align}
i\hbar &  \frac{\partial}{\partial t}\psi(x,p_{y};t)    =\frac{p_{y}^{2}}{2m}%
\psi(x,p_{y};t)-\frac{\hbar^{2}}{2m}\frac{\partial^{2}\psi(x,p_{y}%
;t)}{\partial x^{2}} \nonumber \label{Schxp_y}\\
\nonumber\\
 &  +V\left[  x,i\hbar\left(
-\theta\frac{\partial}{\partial x}+\frac {\partial}{\partial
p_{y}}\right)  \right] \psi(x,p_{y};t).
\end{align}
\bigskip

\noindent Substituting (\ref{Schxp_y}) in this relation one shows
that, at least for potentials which are quadratic in the components of the postition operator, the equation reduces to:
\begin{align}
 \frac{\partial}{\partial t}~W(z,t)
  =\left\{  H,W(z)\right\}  ,\nonumber \label{WLio}\\
\end{align}
which is nothing but the familiar Liouville equation  applied to
the probability distribution function in phase space. \ Even if
the Wigner function does not satisfy all the requirements of an
authentic probability distribution function, it is subject to the
same mathematical relations as a real one. \

\bigskip

\section{QM Examples on the NC plane}

\bigskip

\subsection{Free Particle}

\bigskip

\noindent We begin with the simplest problem, namely the free
particle in two NC dimensions. \ The corresponding Hamiltonian is
\begin{equation}
\hat{H}=\frac{\hat{p}_{x}^{2}+\hat{p}_{y}^{2}}{2m} \nonumber.
\end{equation}
The Schr\"{o}dinger equation in the momentum representation
determines the spectrum of energies to be the continuum
\begin{equation}
\hat{H}\psi(p;t)=i\hbar\frac{\partial}{\partial t}\psi(p;t)=\frac{p_{x}%
^{2}+p_{y}^{2}}{2m}\psi(p;t), \label{Spl} \nonumber%
\end{equation}
\begin{equation}
E=\frac{p_{x}^{2}+p_{y}^{2}}{2m}. \nonumber%
\end{equation}
\noindent The general solution is a superposition of stationary
eigenfunctions of the Hamiltonian:
\begin{equation}
\psi(p;t)=\int a_{E}e^{-\frac{i}{\hbar}E t}\psi_{E}(p)dE,
\end{equation}

\noindent However, (\ref{Spl}) does not provide any information
about the wave functions $\psi(p)$. \ In order to determine these
functions, we will take into account Schr\"{o}dinger equation in
the  $\left| x,p_{y}\right\rangle $ and
$\left|  y,p_{x}\right\rangle $ basis;%
\begin{align}
\hat{H}\psi(x,p_{y})  &
=-\frac{\hbar^{2}}{2m}\frac{\partial^{2}\psi
(x,p_{y})}{\partial x^{2}}+\frac{p_{y}^{2}}{2m}\psi(x,p_{y}),\\
\hat{H}\psi(y,p_{x})  &
=-\frac{\hbar^{2}}{2m}\frac{\partial^{2}\psi (y,p_{x})}{\partial
y^{2}}+\frac{p_{x}^{2}}{2m}\psi(y,p_{x}).\nonumber
\end{align}
Using separation of variables to solve those equations we obtain:
\begin{align}
\psi(x,p_{y})  &  =\frac{e^{\frac{i}{\hbar}xp_{x}}}{\sqrt{2\pi\hbar}}%
\phi(p_{y}),\\
\psi(y,p_{x})  &  =\frac{e^{\frac{i}{\hbar}yp_{y}}}{\sqrt{2\pi\hbar}}%
\varphi(p_{x}).\nonumber
\end{align}
where $\varphi(p_{x})$ and $\phi(p_{y})$ are undetermined
functions. The wave function in momentum space is related to these
by means of the following transforms (\ref{transf}):
\begin{align}
\psi(p^{\prime})  &  =\int_{-\infty}^{\infty} \frac{dydp_{x}}{2\pi\hbar}%
\delta(p_{x}-p_{x}^{\prime})e^{-\frac{i}{\hbar}\left[  yp_{y}^{\prime}%
-\frac{\theta}{2}p_{y}^{\prime}p_{x}^{\prime}\right]
}e^{\frac{i}{\hbar
}yp_{y}}\varphi(p_{x}),\nonumber\\
&
=\varphi(p_{x}^{\prime})e^{\frac{i}{\hbar}\frac{\theta}{2}p_{y}^{\prime
}p_{x}^{\prime}}\delta(p_{y}-p_{y}^{\prime}),\\
\psi(p^{\prime})  &  =\int_{-\infty}^{\infty}\frac{1dxdp_{y}}{2\pi\hbar}%
\delta(p_{y}-p_{y}^{\prime})e^{-\frac{i}{\hbar}\left[  xp_{x}^{\prime}%
+\frac{\theta}{2}p_{y}^{\prime}p_{x}^{\prime}\right]
}e^{\frac{i}{\hbar }xp_{x}}\phi(p_{y}), \\
&  =\phi(p_{y}^{\prime})e^{-\frac{i}{\hbar}\frac{\theta}{2}p_{y}^{\prime}%
p_{x}^{\prime}}\delta(p_{x}-p_{x}^{\prime}).
\end{align}

\noindent Since $\psi(p)$ is by assumption separable, then
\begin{equation}
\psi(p)=\delta(p_{x}-p_{x}^{\prime})\delta(p_{y}-p_{y}^{\prime}).
\end{equation}
\noindent The same result is obtained in the conventional QM,
which is not surprising, for in both cases $\hat{p}_{x}$ and
$\hat{p}_{y}$ commute with the Hamiltonian and therefore have
common eigenfunctions.

\bigskip

\subsection{Harmonic Oscillator}

\bigskip

Next we consider the 2D Isotropic Harmonic Oscillator, described
by the Hamiltonian:
\begin{equation}
\hat{H}=\frac{1}{2m}\left(  \hat{p}_{x}^{2}+\hat{p}_{y}^{2}\right)
+\frac
{1}{2}m\omega^{2}(\hat{x}^{2}+\hat{y}^{2}). \label{hqho}%
\end{equation}
\noindent We will work in the momentum representation of the wave
function. Since the potential does not depends explicitly on time,
the problem reduces to the eigenvalue equation:
\begin{align}
E & \psi(p)   =\frac{1}{2m}(p_{x}^{2}+p_{y}^{2})\psi(p)~+ \\
& + \frac{1}{2}m\omega^{2}\left[  \left(  i\hbar\frac{\partial}{\partial p_{x}%
}-\frac{\theta}{2}p_{y}\right)  ^{2}+\left(
i\hbar\frac{\partial}{\partial p_{y}}+\frac{\theta}{2}p_{x}\right)
^{2}\right]  \psi(p) \nonumber.
\end{align}
Rearranging terms, the last equation can be written as
\begin{align} E\psi(p)  &  =\left(
1+\frac{m^{2}\omega^{2}\theta^{2}}{4}\right) \left\{
\frac{1}{2m}(p_{x}^{2}+p_{y}^{2})\psi(p)\right. \label{qho}\\
&  \left.  -\frac{\hbar^{2}m\omega^{2}}{2\left(  1+\frac{m^{2}\omega^{2}%
\theta^{2}}{4}\right)  }\left[  \frac{\partial^{2}}{\partial p_{x}^{2}}%
+\frac{\partial^{2}}{\partial p_{y}^{2}}\right]  \psi(p)\right\} \nonumber\\
&  -\frac{i}{2}\hbar\theta m\omega^{2}\left(  \frac{\partial}{\partial p_{x}%
}p_{y}-\frac{\partial}{\partial p_{y}}p_{x}\right)
\psi(p).\nonumber
\end{align}
\bigskip

\noindent On the other hand, the Hamiltonian (\ref{hqho}) is
rotationally invariant, that is, it commutes with the quantum
version of the angular momentum Eq.(\ref{MomAng}):
\begin{equation}
\hat{J}=\hat{x}\hat{p}_{y}-\hat{y}\hat{p}_{x}+\frac{\theta}{2}\left(
\hat {p}_{x}^{2}+\hat{p}_{y}^{2}\right).
\end{equation}
\noindent At this point it is convenient to recall that according
to our analysis of the symmetries of the classical NC harmonic
oscillator, SU(2) \textbf{is not} a symmetry for this system.
Thus, for the harmonic oscillator, $\hat{H}$ and $\hat{J}$ have
common eigenstates. \ \ Noticing that the angular momentum
operates over the Harmonic oscillator eigenfunctions as follows:
\begin{equation}
\hat{J}\psi(p)=i\hbar\left(  \frac{\partial}{\partial p_{x}}p_{y}%
-\frac{\partial}{\partial p_{y}}p_{x}\right)  \psi(p).
\end{equation}
\noindent the eigenvalue equation for the Hamiltonian $\hat{H}$
can be written in terms of $\hat{J}$ as:
\begin{equation}
E\psi(p)=\left(  1+\frac{m^{2}\omega^{2}\theta^{2}}{4}\right)  \hat{H}_{0}%
\psi(p)-\frac{1}{2}\theta m\omega^{2}\hat{J}\psi(p),
\end{equation}
\noindent where we introduced the operator
\begin{equation}
\hat{H}_{0}=\left\{ \frac{1}{2m}(p_{x}^{2}+p_{y}^{2})-\frac{\hbar
^{2}m\omega^{2}}{2\left(
1+\frac{m^{2}\omega^{2}\theta^{2}}{4}\right) }\left[
\frac{\partial^{2}}{\partial
p_{x}^{2}}+\frac{\partial^{2}}{\partial
p_{x}^{2}}\right]  \right\}, \nonumber \label{Hcero}%
\end{equation}
which is nothing but the Hamiltonian of a commutative harmonic
oscillator of frequency $\varpi$:
\begin{equation}
\varpi=\frac{\omega}{\sqrt{1+\frac{m^{2}\omega^{2}\theta^{2}}{4}}}.
\end{equation}

\noindent Besides $\left[  \hat{H},\hat{J}\right]  =0$, one can
shown that $\left[ \hat{H}_{0},\hat{J}\right]  =0$ and $\left[
\hat{H},\hat{H}_{0}\right]  =0$ hold. \ Therefore $\hat{H}$, $\hat
{H}_{0}$ and $\hat{J}$ have common eigenstates . The solution to
the partial differential equation (\ref{Hcero}) is the product of
harmonic oscillator eigenfunctions:
\begin{equation}
\psi_{n_{1},n_{2}}(p)=\frac{e^{-\frac{p_{x}^{2}+p_{y}^{2}}{2m\hbar\varpi}}%
}{N}H_{n_{1}}%
(\frac{p_{x}}{m\hbar\varpi})H_{n_{2}}(\frac{p_{y}}{m\hbar\varpi}),
\end{equation}
where $N=2^{n_{1}+n_{2}}\sqrt{\pi
m\hbar\varpi}\sqrt{n_{1}!n_{2}!}.$  Thus, the eigenvalue equation
for $\hat{H}_{0}$ is: \
\begin{equation}
\hat{H}_{0}\psi_{n_{1},n_{2}}(p)=\hbar\varpi\left(
n_{1}+n_{2}+1\right) \psi_{n_{1},n_{2}}(p),
\end{equation}
with $n_{1},n_{2}$ positive integers. \ These functions form a
complete orthonormal basis. \ The lineal combination of these
functions that  simultaneously diagonalizes $\hat{H}$ and
$\hat{J}$ ~is:

\begin{align}
&\psi_{n,j}(p)    =\frac{e^{-\frac{p_{x}^{2}+p_{y}^{2}}{2m\hbar\varpi}}}%
{\bar N}\sum_{r=0}^{\frac{n}{2}+j}\sum_{q=0}^{\frac{n}{2}%
-j}\binom{\frac{n}{2}+j}{r}\times  \binom{\frac{n}{2}-j}{q} \label{HoEf}\\
& \times(-1)^{q}(i)^{r+q}H_{n-(r+q)}\left(
\frac{p_{x}}{m\hbar\varpi}\right)  H_{r+q}\left(
\frac{p_{y}}{m\hbar\varpi }\right)  ,\nonumber
\end{align}

\noindent with $n\in\mathbb{Z}^{+}\ $,  $2j\in\mathbb{Z}$,
subjects to the restriction $-\frac{n}{2}\leq j\leq\frac{n}{2}$
and $\bar N = \ 2^{n}\sqrt{\pi m\hbar\varpi}\sqrt{\left(
\frac{n}{2}+j\right) !\left( \frac{n}{2}-j\right)  !}.$  Putting
all together the eigenvalue spectrum associated to eigenfunctions
(\ref{HoEf}) is summarized in the equations:
\begin{align}
\hat{H} & \psi_{n,j}(p)    =E_{n,j}\psi_{n,j}(p)  \\
&  =\left[
\hbar\omega\sqrt{1+\frac{m^{2}\omega^{2}\theta^{2}}{4}}\left(
n+1\right)  -\theta m\omega^{2}\hbar j\right]
\psi_{n,j}(p),\nonumber
\end{align}%
\begin{equation}
\hat{J}\psi_{n,j}(p)=2\hbar j\psi_{n,j}(p) \nonumber.
\end{equation}

\bigskip

\noindent To conclude the discussion on the harmonic oscillator,
we derive the Wigner quasi-distribution  function for the ground
state of the harmonic oscillator, which is described by the wave
functions
\begin{equation}
\psi_{0,0}(p)=\frac{e^{-\frac{p_{x}^{2}+p_{y}^{2}}{2m\hbar\varpi}}}{\sqrt{\pi
m\hbar\varpi}}.
\end{equation}
The Wigner distribution function is, according to Eq.(\ref{Wp}):
\begin{align}
W & _{0,0}(z)
=\frac{e^{-\frac{p_{x}^{2}+p_{y}^{2}}{m\hbar\varpi}}}{\sqrt
{\pi^{5}m\hbar^{5}\varpi}}\times \nonumber\\
&  \times\int_{-\infty}^{\infty}dud\eta e^{-\frac{u^{2}+\eta^{2}}%
{2m\hbar\varpi}}e^{-\frac{2i}{\hbar}\left[  xu+y\eta-\frac{\theta}{2}%
(p_{x}\eta-p_{y}u)\right]  },\nonumber\\
&=\frac{1}{\pi^{2}\hbar^{2}}e^{\left\{  -\frac{p_{x}^{2}+p_{y}^{2}%
}{m\hbar\varpi}-\frac{m\varpi}{\hbar}\left[  \left(  x+\frac{\theta}{2}%
p_{y}\right)  ^{2}+\left(  y-\frac{\theta}{2}p_{x}\right)
^{2}\right] \right\}  }.
\end{align}

\noindent This function is positive definite in its entire domain.
Without loss of generality let us consider for simplicity the case
$m=1,\hbar=1,\omega=1$ (which can be obtained  through a scale
transformation). \ The time evolution of Wigner function is
determined by Eq.(\ref{WLio}). The solution to this equation can
be written operationally as:
\begin{equation}
W_{0,0}(z,t)=\exp\left[  it \left\{  H,\cdot\right\}  \right]
W_{0,0}(z,0),
\end{equation}
which is nothing but a linear, time dependent transformation.
Then, the complete solution is
\begin{align}
W_{0,0}(z,t) &  =\exp\left[  it \left\{  H,\cdot\right\}  \right]  \times\\
&  \times\frac{1}{\pi^{2}\hbar^{2}}\exp\left\{  -\frac{\sqrt{4+\theta^{2}}}%
{2}\left(  p_{x}^{2}+p_{y}^{2}\right)  \right\}  \times\nonumber\\
&  \times\exp\left\{  -\frac{2\left[  \left(
x+\frac{\theta}{2}p_{y}\right) ^{2}+\left(
y-\frac{\theta}{2}p_{x}\right)  ^{2}\right]  }{\sqrt{4+\theta
^{2}}}\right\}  .\nonumber
\end{align}

\bigskip

\subsection{2D Solid (Einstein's model)}
\bigskip

\noindent The thermodynamic properties of this solid can be
studied considering a canonical ensemble of $N$ distinguishable,
non interacting and NC harmonic oscillators \cite{LandSP}. \ For
an oscillator of frequency $\omega$, the probability of being in
the $n,j$ state, denoted $w_{n,j}$, is
\begin{equation}
w_{n,j}=\frac{e^{-\frac{E_{n,j}}{kT}}}{Z(T,V,1)},
\end{equation}
where $Z(T,V,1)$ is the partition function
\begin{equation}
Z(T,V,1)=Tr(e^{-\frac{\hat{H}}{kT}})=\sum_{n=0}^{\infty}\sum_{j=-n/2}%
^{n/2}e^{-\frac{E_{n,j}}{kT}}.
\end{equation}
The calculation of this function proceeds through geometric sums,
and renders the following result:
\begin{align}
&  Z(T,V,1)\\
&  =\left\{  2\left[  \cos\left(
\frac{\hbar\omega\Theta}{2kT}\right) -\cos\left(
\frac{\hbar\omega^{2}m\theta}{2kT}\right)  \right]  \right\}
^{-1},\nonumber
\end{align}
with
\begin{equation}
\Theta=\sqrt{4+m^{2}\omega^{2}\theta^{2}}.
\end{equation}
At this point, it is necessary to know the distribution function
of the natural frequencies in the solid. To simplify the
calculation we choose Einstein's approximation, and set all
frequencies equal $\omega_{i}=\omega$. This approach, if not the
better, provides a clear qualitative behavior of the system.
\bigskip

\noindent The partition function of the $N$ oscillators is, then,
\begin{equation}
Z(T,V,N)=Z(T,V,1)^{N}.
\end{equation}
{From} this partition function, the derivation of the Free Energy is
immediate
\begin{align}
A(T,V,N)  &  =-kT\ln\left[  Z(T,V,N)\right]  ,\\
&  =NkT\ln\left[  \cos\left(  \frac{\hbar\omega\Theta}{2kT}\right)
-\cos\left(  \frac{\hbar\omega^{2}m\theta}{2kT}\right)  \right] \nonumber\\
&  +NkT\ln2.\nonumber
\end{align}
Thus, the entropy of the system is
\begin{align}
S  &  =-\left.  \frac{\partial A}{\partial T}\right|
_{V,N}=-Nk\ln\left[ \cos\left(
\frac{\hbar\omega\Theta}{2kT}\right)  -\cos\left(  \frac
{\hbar\omega^{2}m\theta}{2kT}\right)  \right]\nonumber \label{ent0}\\
&  +\frac{\hbar\omega N\left[
\sqrt{4+m^{2}\omega^{2}\theta^{2}}\sin\left(
\frac{\hbar\omega\Theta}{2kT}\right)  -m\omega\theta\sin\left( \frac
{\hbar\omega^{2}m\theta}{2kT}\right)  \right]  }{2T\left[ \cos\left(
\frac{\hbar\omega\Theta}{2kT}\right)  -\cos\left( \frac{\hbar\omega
^{2}m\theta}{2kT}\right)  \right]  }\nonumber\\
&  -Nk\ln2,
\end{align}
and its internal energy takes the form
\begin{align}
U  &  =A+TS\\
&  =\frac{\hbar\omega N\left[  \Theta\sin\left(  \frac{\hbar\omega\Theta}%
{2kT}\right)  -m\omega\theta\sin\left(  \frac{\hbar\omega^{2}m\theta}%
{2kT}\right)  \right]  }{2T\left[  \cos\left(  \frac{\hbar\omega\Theta}%
{2kT}\right)  -\cos\left(
\frac{\hbar\omega^{2}m\theta}{2kT}\right)  \right] }.\nonumber
\end{align}

\bigskip

\noindent In the high temperature regime $kT\gg\hbar\omega$, the
behavior of $U$ can be deduced expanding in power series of $T:$%
\begin{equation}
U_{kT\gg\hbar\omega}=2NkT+\frac{\hbar^{2}\omega^{2}N\left(
2+m^{2}\omega ^{2}\theta^{2}\right)  }{12kT}+\cdots,
\end{equation}
\noindent We conclude that the conventional energy equipartition
is also obtained in the NC case. \ On the other hand, in the
opposite limit $T\rightarrow0$, \ the internal energy reduces to
the contribution of the minimum energy of each oscillator
\begin{equation}
\lim_{T\rightarrow0}U=N\hbar\omega\sqrt{1+\frac{m^{2}\omega^{2}\theta^{2}}{4}%
}.
\end{equation}

\bigskip

\noindent Finally, the calorific capacity of this set of NC
oscillators is given by
\begin{align}
C_{V}  &  =\left.  \frac{\partial U}{\partial T}\right|  _{V,N}\\
&  =\frac{\hbar^{2}\omega^{2}N}{2kT^{2}}\left[  \cos\left(
\frac{\hbar
\omega\Theta}{2kT}\right)  -\cos\left(  \frac{\hbar\omega^{2}m\theta}%
{2kT}\right)  \right]  ^{-2}\times\nonumber\\
&  \left\{  \left(  2+m^{2}\omega^{2}\theta^{2}\right)  \left[
\cos\left(
\frac{\hbar\omega\Theta}{2kT}\right)  \cos\left(  \frac{\hbar\omega^{2}%
m\theta}{2kT}\right)  -1\right]  \right. \nonumber\\
&  \left.  -m\omega\theta\Theta\left[  \sin\left(
\frac{\hbar\omega\Theta }{2kT}\right)  \sin\left(
\frac{\hbar\omega^{2}m\theta}{2kT}\right)  \right] \right\}
.\nonumber
\end{align}
Again, in the high temperature regime one recovers the
conventional result:
\begin{equation}
C_{V}=2kN-\frac{\hbar^{2}\omega^{2}N\left(
2+m^{2}\omega^{2}\theta ^{2}\right)  }{12kT^{2}}+\cdots.
\end{equation}

\bigskip

\begin{figure}[h]
\includegraphics[width=8cm]{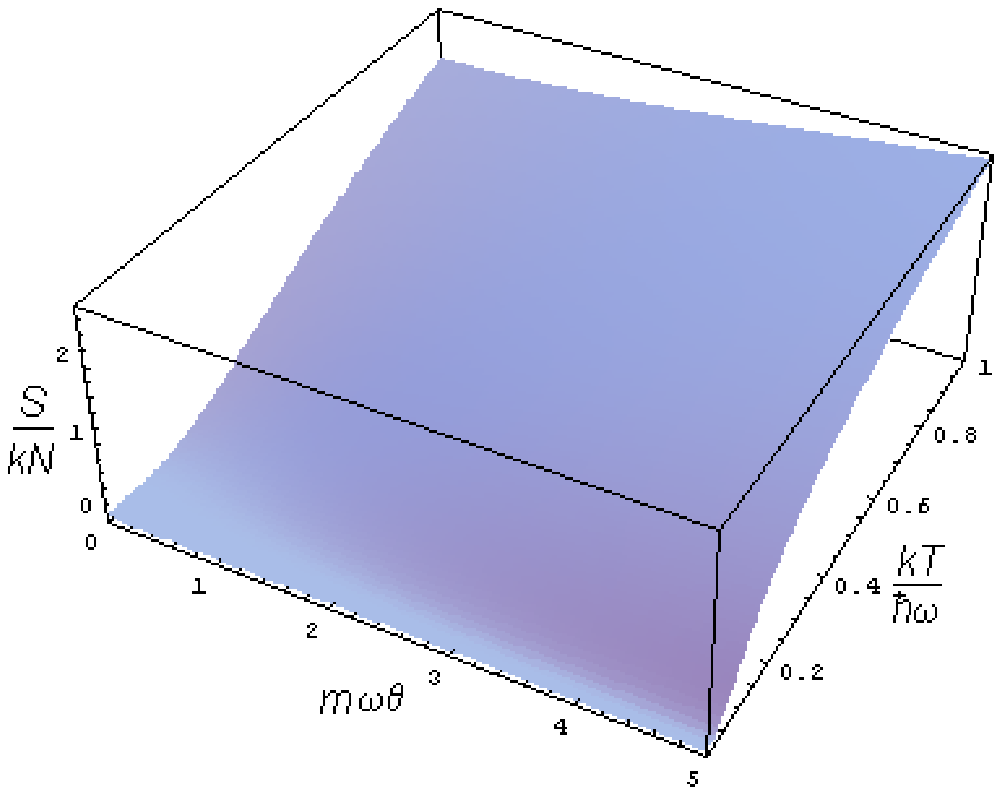}
 \caption{Figure 1}\label{ent1}
\end{figure}

\begin{figure}[h]
\includegraphics[width=8cm]{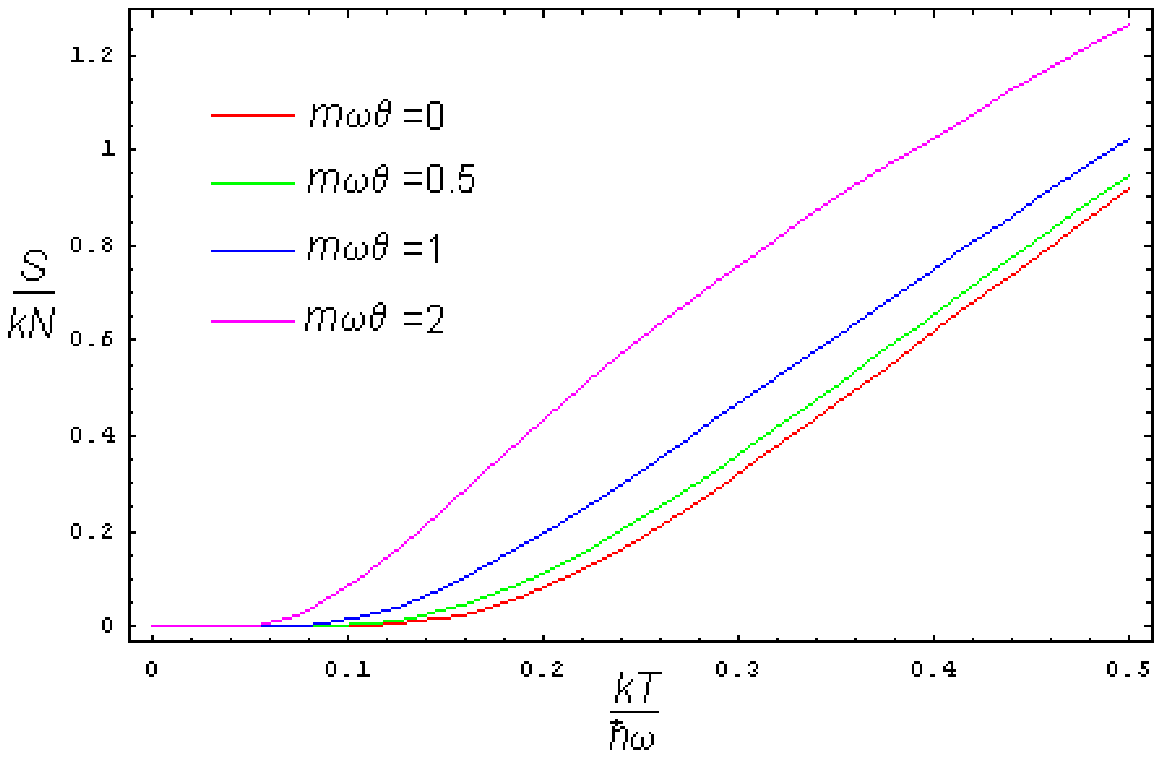}
\caption{Figure 2}\label{ent2}
\end{figure}

\noindent Finally in figures  \ref{ent1} and \ref{ent2} we show
the behavior of the entropy $S$ as a function of $T$ and $\theta$,
for fixed values of $m,\omega$ and $N$. In the present model, the
entropy varies significatively with respect to $\theta$, in
particular if ~$\theta \sim (m\omega)^{-1}$.%

\bigskip

\section{summary and conclusions}
\bigskip

\noindent We presented a systematic study of non-commutative
mechanics starting from the classical formalism and proceeding
through the quantization. We emphasized the role played by
symmetries, in particular we ensure that the NC free particle is
consistent with Galilean relativity which follows from the well
known relation among the boost generators and the position
operator. The general description of NCCM in terms of second class
constrained system was elaborated for Hamiltonians of the type $H =
T + V$ where $T$ and $V$ stand for kinetic and potential energy of
the two dimensional system.
\bigskip

\noindent Besides providing a global view of the problem, our
manuscript contains new results, in particular:
\bigskip

\begin{itemize}
\item A formulation that avoids the use of  non-canonical
transformations and/or expansion in the NC parameter $\theta$.
\item In classical mechanics, analytical solution for the free
particle and harmonic oscillator problems. We also show that $SU(2)$
is not the symmetry group of the isotropic harmonic oscillator, and
identify the generators of the existing symmetry.

\item Quantization is presented in three different frameworks:
Canonical, Path Integral and Wigner Function.
\bigskip

\item The
representations of the Heisenberg Algebra in three different basis
($(p_x,p_y)$, $(x,p_y)$ and $(p_x,y)$), including the wave functions
permitting the change of basis. \item Non-equivalent representations
of the Heisenberg algebra  characterized by gauge fields that follow
naturally from the structure of the algebra. Those Fields are
relevant in the description of multiply connected manifolds. \item
Extension of the analysis of the fundamental properties of the
Wigner function in four dimensional NC phase space. \item Analytical
solution in QM for the free particle and harmonic oscillator without
performing the customary non-canonical transformation, without using
the structure of $SU(2)$ generators and without assuming  \textsl{a
priori} the existence of a vacuum state. \item The thermodynamic
properties of a 2D NC crystal using the 2D Einstein's solid model.
The behavior of the entropy as a function of the NC parameter
$\theta$ is reported.
\end{itemize}
\noindent The proper definition of the system to be treated in NC
mechanics and the certitude that it is not traded  along the way are
fundamental. Our approach focuses on both points, first defining a
general consistent framework and second avoiding completely the use
of non canonical transformations since those lead, in general, to a
system whose properties are completely different from those of the
original one.

\begin{acknowledgments}
Work partly supported by CONACyT under grant 37234-E, CONCyTEG
04-16-K117-027 and DINPO-UGTO. The authors gratefully acknowledge
financial support from ICTP, where this work was concluded.
\end{acknowledgments}

\newpage


FIGURE CAPTION.

Figure 1. Normalized entropy as a function of the NC parameter
$\theta$ and temperature as obtained from Eq.(\ref{ent0}).

Figure 2. Normalized entropy as a function of temperature, for fixed
values of the NC parameter $\theta$ as obtained from
Eq.(\ref{ent0}).

\newpage

%$^1$ In the following we will use NC to stand for
%non-commutative and M, C and Q for mechanics classical and quantum
%respectively.

%$^2$ At this point it is not clear the relation between the
%Galilei algebra and the symplectic structure. The connection
%between these structures is seen when the generators of the
%Galilei group are expressed in terms of the dynamical variables of
%a physical system, see below.

%$^3$ The $\hbar$ enters naturally in the quantization procedure,
%just on dimensional grounds one requires it. In the NC case the
%situation is different, in fact one should keep in mind the
%appropriated units of $\theta$, which are not the same that at the
%classical level.

\newpage

%\simple


\begin{thebibliography}{}

    \bibitem{connes}
    A. Connes, Non commutative Geometry, Academic Press, San
    Diego (1994).

\bigskip
    \bibitem{witten}
    N. Seiberg, E. Witten, JHEP 0002 (2000) 020.

\bigskip
    \bibitem{luki}
    J.Lukierski, P.C. Stichel W.J. Zakrzewski,
    Annals Phys 260 (1997) 224.

\bigskip
    \bibitem{levy}
    J.-M. L\'evi-Leblond, Riv. Nuovo Cimento 4,1 (1974) 99.

\bigskip
    \bibitem{janb}
    G. Dunne, R. Jackiw, Nucl. Phys. Proc. Suppl. 33C
    (1993) 114, hep-th/9204057; C. Duval, P. A. Horváthy, Phys.Lett. B479
    (2000) 284 , hep-th/0002233.

\bigskip
    \bibitem{janbb}
    J. Govaerts, Hamiltonian Quantisation and Constrainded
    Dynamics, Leuven University Press, Leuven, 1991.

\bigskip
    \bibitem{jackiw}
    See G. Dune, Jackiw in Ref(\cite{janb}) and also A.A. Deriglasov, Noncommutative
    version of an arbitrary nondegenerate mechanics,
    hep-th/0208072.

\bigskip
    \bibitem{free}
    S. Chaturvedi, R. Jagannathan, R. Sridhar, V. Srinivasan, J.
    Phys. A: Math. Gen. 26 L105-L112.

\bigskip
    \bibitem{lumbo}
    Musongela Lubo JHEP 0405 (2004) 045,  hep-th/0304039.

\bigskip
    \bibitem{polyn}
    V.P. Nair, Polychronakos, Phys. Lett. B 505 (2001) 267,
    hep-th/0011172; see also Kang Li,
    J. Wang, C. Chen, Representation of Noncommutative phase space,
    Hep-th/0407183.

\bigskip
    \bibitem{brihaye}
    Y. Brihaye, C. Gonera, S. Giller, P. Kosinski, Galilean invariance in 2+1
    dimensions, hep-th/9503046.

\bigskip
    \bibitem{janv}
    J. Govaerts, V. Villanueva, Int. J. Mod. Phys. A15 (2000) 4903, quant-ph/9908014

\bigskip
    \bibitem{aci}
     C. Acatrinei, Comments on noncommutative particle dynamics, hep-th/0106141

\bigskip
    \bibitem{op1}
    See for example J.S. Bell, Speakable and Unspeakable in Quantum
    Mechanics, Cambridge University press, Cambridge UK, 1987; K.
    Banaszek, K. Wodkiewicz, Phys. Rev. A 58 (1998) 4345.

\bigskip
    \bibitem{koka}
    A. Kokado, T. Okamura, T. Saito, Wigner's formulation of Noncommutative Quantum
    Mechanics, hep-th/0208040; O.F. Dayi, L.T. Kelleyane, Mod.Phys.Lett. A17
    (2002)1937 hep-th/0202062; M. Rosenbaum, J.D. Vergara, The *-value equation and
    Wigner distribution in noncommutative Heisenberg algebras, hep-th/0505127.

\bigskip
    \bibitem{osc1}
    S. Bellucci, A. Nersessian and C. Sochichiu,  Phys.Lett. \textbf{B522}: 345
    (2001), hep-th/0106138

\bigskip
    \bibitem{osc2}
    A. Smailagic, E. Spallucci, Phys. Rev. D65 (2002) 107701; I. Dadic, L. Jonke,
    S. Meljanac, Harmonic oscillator on noncommutative spaces, hep-th/0301066.

\bigskip
    \bibitem{su(2)}
    J. Gamboa, M. Loewe, C. Rojas, Int. J. Mod. Phys. A17 (2002) 2555;
    H.O. Girotti, Am. J. Phys. 72 (2004) 608.

\bigskip
    \bibitem{?gold}
    H. Goldstein, Clasical Mechanics, Addison Wesley, Reading, Massachusetts
    1980.

\bigskip
    \bibitem{gal}
    D. R. Grigore,  Journ. Math. Phys. 34 (1993) 4190, hep-th/9312048.

\bigskip
    \bibitem{classical}
    Juan M. Romero, J.A. Santiago, J. David Vergara, Phys. Lett. A310
    (2003) 9, hep-th/0211165; Juan M. Romero, J.David Vergara, Mod. Phys. Lett. A18
    (2003) 1673, hep-th/0303064; A.E.F. Djemai, On noncommutative classical
    mechanics, hep-th/0309034.

\bigskip
    \bibitem{cvtesis}
    C. Vaquera-Araujo, No conmutatividad en 2 Dimensiones, Bachelor thesis,
    Instituto de Fisica, Universidad de Guanajuato (2005), Unpublished.

\bigskip
    \bibitem{symmetries}
    O. Espinoza, P. Gaete, Symmetries in noncommutative quantum
    mechanics, hep-th/0206066.
\bigskip    
   \bibitem{quien}
    B. DeWitt, Rev. Mod. Phys. \textbf{29} (1957) 377

\bigskip
    \bibitem{Dirac}
    P. A. M. Dirac, Lectures on Quantum Mechanics, Belfer
    Graduate School of Science, Yeshiva University, New York,
    1964.

\bigskip
    \bibitem{gov}
    J. Govaerts, V. Villanueva, Int. J. Mod. Phys A15 (2000)
    4903, quant-ph/9908014.

\bigskip
    \bibitem{Wigner1}
    E. Wigner, Phys. Rev. 40 (1932) 749.

\bigskip
     \bibitem{LandSP}
     L.D. Landau, Butterwoth, Heinemann, Statistical Physics, Course of theoretical
     Physics, Vol. 5, 2000.

\end{thebibliography}
\end{document}